\documentclass[aps,prb,a4,floatfix,amssymb,amsfonts,amsmath,reprint,superscriptaddress]{revtex4-2}

\usepackage{graphicx}
\usepackage{amsmath}
\usepackage{dcolumn}
\usepackage{bm}
\usepackage{textcomp}

\usepackage{xcolor}

\def\cm{cm$^{-1}$}
\def\TF{TCNQF$_4$}

\begin{document}

\title{Tetramethylbenzidine-TetrafluoroTCNQ (TMB-\TF): A narrow-gap semiconducting salt with room temperature relaxor ferroelectric behavior}

\author{Stefano Canossa}
\affiliation{EMAT, Department of Physics, University of Antwerp, 2020 Antwerp, Belgium}
	
\author{Elena Ferrari}
\affiliation{Dipartimento di Scienze Chimiche, della Vita e della Sostenibilità
	Ambientale (S.C.V.S.A.) \& INSTM-UdR Parma, Universit\`a di Parma, 43124 Parma, Italy}

\author{Pit Sippel}
\affiliation{Experimental Physics V, Center for Electronic Correlations and Magnetism, University of Augsburg, 86159 Augsburg, Germany}

\author{Jonas K. H. Fischer}
\affiliation{Experimental Physics V, Center for Electronic Correlations and Magnetism, University of Augsburg, 86159 Augsburg, Germany}
\affiliation{Tohoku Forum for Creativity, Tohoku University, 980-8577 Sendai, Japan}

\author{Raphael Pfattner}
\affiliation{Department of Molecular Nanoscience and Organic Materials, Institut de Ci\`encia de Materials de Barcelona (ICMAB-CSIC) and Networking Research Center on Bioengineering, Biomaterials and Nanomedicine (CIBER-BBN),  ES-08193 Bellaterra, Spain}

\author{Ruggero Frison}
\affiliation{Physik-Institut, Universit\"at Z\"urich, 8057 Z\"urich, Switzerland}

\author{Matteo Masino}
\affiliation{Dipartimento di Scienze Chimiche, della Vita e della Sostenibilità
	Ambientale (S.C.V.S.A.) \& INSTM-UdR Parma, Universit\`a di Parma, 43124 Parma, Italy}

\author{Marta Mas-Torrent}
\affiliation{Department of Molecular Nanoscience and Organic Materials, Institut de Ci\`encia de Materials de Barcelona (ICMAB-CSIC) and Networking Research Center on Bioengineering, Biomaterials and Nanomedicine (CIBER-BBN),  ES-08193 Bellaterra, Spain}

\author{Peter Lunkenheimer}
\affiliation{Experimental Physics V, Center for Electronic Correlations and Magnetism, University of Augsburg, 86159 Augsburg, Germany}

\author{Concepci\'o Rovira}
\affiliation{Department of Molecular Nanoscience and Organic Materials, Institut de Ci\`encia de Materials de Barcelona (ICMAB-CSIC) and Networking Research Center on Bioengineering, Biomaterials and Nanomedicine (CIBER-BBN),  ES-08193 Bellaterra, Spain}

\author{Alberto Girlando$^*$}
\affiliation{Dipartimento di Scienze Chimiche, della Vita e della Sostenibilità
Ambientale (S.C.V.S.A.) \& INSTM-UdR Parma, Universit\`a di Parma, 43124 Parma, Italy}
\affiliation{Present address: Molecular Materials Group, 43124 Parma, Italy}

\begin{abstract}
We present an extension and revision of the spectroscopic and structural data
of the mixed stack charge transfer (CT) crystal 3,3$^\prime$,5,5$^\prime$-tetramethylbenzidine--tetrafluoro-tetracyanoquinodimethane (TMB-\TF),
associated with new electric and dielectric measurements. Refinement of syncrotron structural
data at low temperature has led to revise the previously reported [Phys. Rev. Mat. \textbf{2}, 024602 (2018)] $C2/m$ structure. The revised structure is $P2_1/m$, with two dimerized
stacks per unit cell, and is consistent with the vibrational data. However, polarized Raman
data in the low-frequency region also indicate that by increasing temperature above 200 K the structure presents an increasing degree of disorder mainly along the stack axis.
X-ray diffraction data at room temperature have confirmed that the correct structure
is $P2_1/m$ - no phase transitions - but did not allow to definitely substantiate the presence of disorder. 
On the other hand, dielectric measurement have evidenced a typical relaxor ferroelectric behavior already at room temperature, with a peak in real part of dielectric constant $\epsilon'(T,\nu)$ around 200 K and 0.1 Hz. The relaxor behavior is explained in terms of the presence of spin solitons separating domains of opposite polarity that yield to ferroelectric nanodomains. 
TMB-\TF~ is confirmed to be a narrow gap band semiconductor ($E_a \sim 0.3$ eV) with room temperature conductivity of $\sim 10^{-4}~ \Omega^{-1}$ \cm.   
\end{abstract}

\maketitle

\section{Introduction}

Over the last few years, mixed-stack (ms) organic charge-transfer (CT)
crystals have attracted great interest due to their unique properties of high tunability and promising applications in several fields of organic electronics,
from Organic Field Effect Transistors (OFET) \cite{Zhang2018, Salzillo2019}, to photovoltaic devices \cite{Kang2013},
to organic ferroelectrics \cite{Horiuchi2014}, and so on. In the case of binary molecular systems, ms-CT crystals consist of a regular arrangement of face-to-face stacks of $\pi-$electron donor (D) and acceptor (A) molecular moieties with a defined stoichiometry.
Such stacks are called mixed to distinguish them from segregated stacks of cation (D$^+$) or anion (A$^-$) radicals in other systems. The ionicity parameter
$\rho$ represents the degree of mixing between ground and excited CT state,
and plays a fundamental role in the physical properties of ms-CT crystals.
The most common and most studied class of 1:1 ms-CT crystals have a neutral or quasi-neutral ground state, $ \rho \lesssim $ 0.5. A handful have intermediate ionicity,
and some of them upon lowering temperature or increasing
pressure undergo a valence instability dubbed Neutral-Ionic transition \cite{Masino2017}. Finally, the strongest electron Donor and Acceptor molecules
form radical ion salts with $\rho \approx 1$, whose Madelung energy
$M$ exceeds the energy cost, $I-A$, to transfer an electron from D to A.

Mixed stack radical salts are rare, also because strong D and A molecules may prefer to crystallize as segregated stacks,
as for instance DBTTF-\TF, \cite{Emge1982} or TMPD-\TF \cite{Meneghetti1988}.
As a matter of fact, old literature search provided just four
ionic ($\rho \geq 0.9$) ms-CT crystals, namely,
TMPD-TCNQ \cite{Girlando1984},  M$_2$P-\TF \cite{Meneghetti1985}, TTF-BA \cite{Girlando1985}, and BEDO-TCNQCl$_2$ \cite{Hasegawa2000}. The three
former all exhibit a spin Peierls transition 
around 200 K, 120 K and 50 K, respectively, whereas BEDO-TCNQCl$_2$ displays
a first order transition around 100-120 K of unclear origin, but without
stack dimerization.

More recently, some of us obtained another ionic ($\rho \simeq 0.9$) ms-CT crystal, TMB-\TF, whose room $T$ vibrational spectra 
clearly indicate a dimerized stack, whereas for X-ray the stack appeared as regular
(equal distances between D and A) \cite{Castagnetti2018}. We have decided to
further investigate this system, also including electric and
dielectric measurements. Indeed, very little is known about semiconducting
properties of ionic ms-CT crystal. Furthermore, the above mentioned
TTF-BA has been the first ms-CT crystal displaying ferroelectric properties in the low-$T$ phase,
where the stack is dimerized \cite{Kagawa2010a}. The new measurements reveal an
actual dimerized stack structure already at room temperature $T$, with an intriguing relaxor
ferroelectric behavior.

\section{Methods}

\subsection{Sample preparation}
The crystals have been prepared by sublimation at about 180 \textdegree C in an open tube under controlled atmosphere \cite{Castagnetti2018}, with a simplified version of the physical transport apparatus described by Laudise \textit{et al.}\cite{Laudise1998}.

\subsection{Electric and dielectric measurements}
Homogeneous thin single crystals of TMB-TCNQF$_4$ were analyzed under an optical microscope equipped with a polarizer/analyzer setup and electrostatically transferred Si/SiO$_2$ substrates with 200 nm thermally grown oxide thickness. Single crystals were electrically connected employing high conductive graphite paste (Dotite XC-12) and thin gold wires at the opposite tip of the needle-like crystals. Crystal dimensions were estimated using an optical microscope. Samples were prepared and electrically connected under ambient conditions (Relative humidity rH=40-60 \% and temperature T = 28$^o$C). All electrical characterization was carried out in darkness within an nitrogen filled glove box with low humidity and oxygen levels (H$_2$0 $<$ 2 ppm, and O$_2$ $<$ 2 ppm) using a Keithley SourceMeter model 2612. Pseudo AC measurements at low current were used to prevent Joule heating of the sample.

The dielectric constant and conductivity were determined using a frequency-response analyzer (Novocontrol Alpha-A). Gold wires were attached to contacts of 
gold paint on opposite tips of the needle-like crystals, ensuring an electric-field direction exactly parallel to the stack axis. Sample cooling and heating were achieved by  
a nitrogen-gas cryosystem (Novocontrol Quatro).

\subsection{Spectroscopic measurements}
Polarized infrared (IR) absorption spectra of the crystals were recorded with a
Bruker IFS-66 Fourier transform spectrometer coupled to the Hyperion 1000 IR microscope  equipped with a wire-grid polarizer. 
Spectral resolution: 2 \cm. The Raman spectra with 752 nm excitation (Lexel Kr laser)
were recorded with a Renishaw 1000 Raman spectrometer
with the appropriate edge filter and coupled to a Leica
M microscope. Raman spactra with the other exciting lines were obtained with 
Horiba LabRAM HR Evolution spectrometer equipped either with a Ultra Low-Frequency (ULF)
Bragg filter or with the appropriate edge filter. Spectral resolution 2 \cm.
Incident and scattered polarization was controlled by a half-wave plate and a thin-film linear polarizer, respectively, and the sample was rotated to record the different polarizations.
A small liquid nitrogen cryostat (Linkam HFS 91) was used for
temperature-dependent measurements under the IR and Raman microscopes.


\subsection{Structural measurements}
Low temperature (100, 150 and 200 K) diffraction data were obtained at the XRD1 beamline
of the Elettra Synchrotron facility (CNR Trieste, Italy), with beam energy of 17.712 keV. Diffraction data at 300 K were collected using a Rigaku-Oxford SuperNova diffractometer,
equipped with Cu K$\alpha$ (8.04 keV) X-ray source. Data collection, refinement procedures and structural analysis are described in detail in the Supporting Information. The structures
have been deposited at the Cambridge Crystallographic Data Center (deposition numbers CCDC 2097174-2097177) and can be obtained
free of charge from the CCDC at www.ccdc.cam.ac.uk/getstructures.


\section{Results}

\subsection{Electric and dielectric measurements}

\begin{figure}[b]
	\centering
	\includegraphics[width=0.8\linewidth]{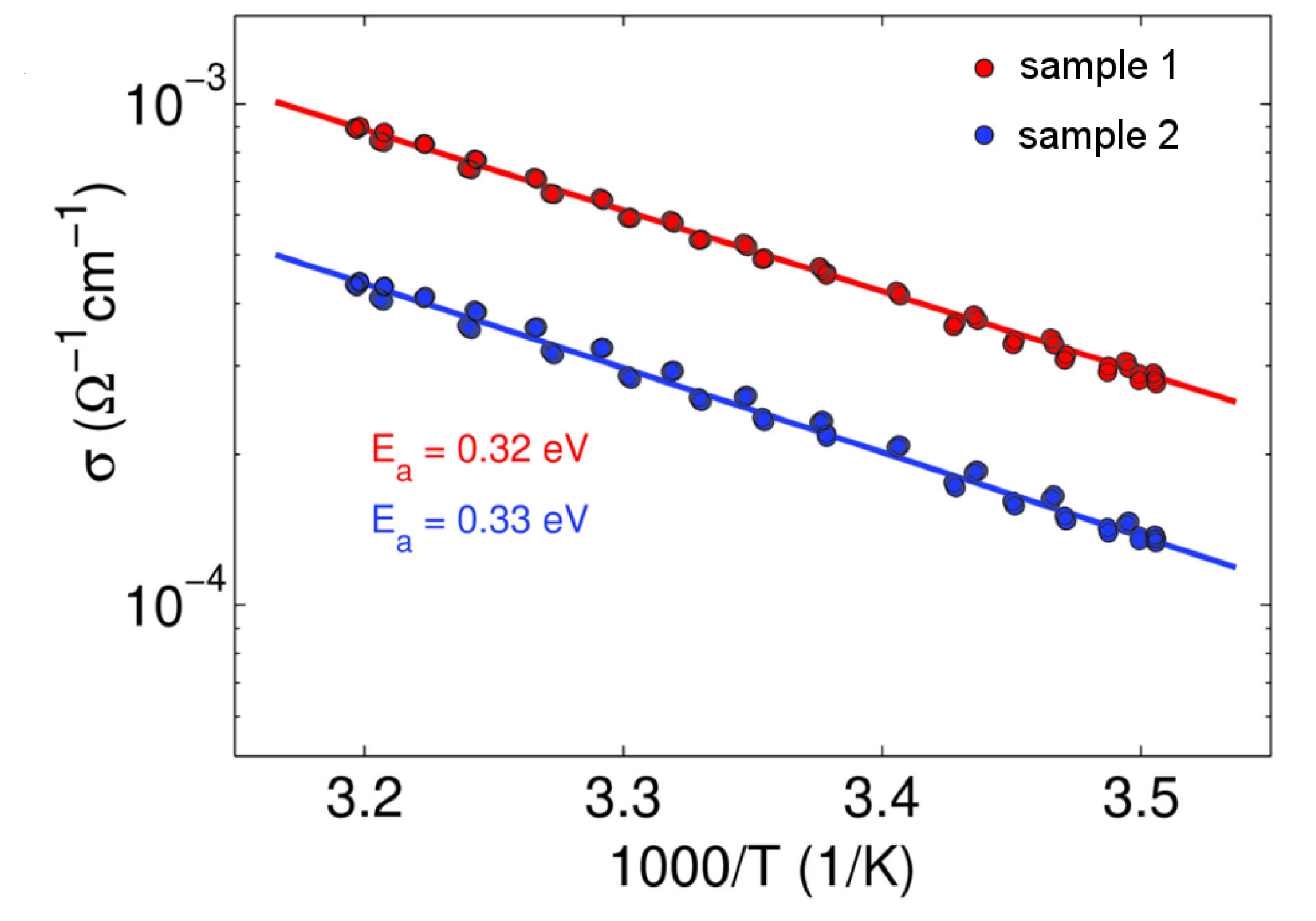}
	\caption{Temperature dependence of the TMB-\TF\ dc-conductivity obtained
		through transistor measurements.}
	\label{fig:tmb-tqf4sigmadc}
\end{figure}

As TMB-\TF~is a rare ionic mixed stack, we believed it was
important to have a characterization of its electric
and dielectric properties, by building transistors and condensers,
as detailed  in the Methods Section and in the Supporting Information. Two crystals, sample 1, Fig. S3 left, and  sample 2, Fig. S3 right, were measured and showed high source-drain currents ($I_D$) in FET measurements at low drain voltages with corresponding gate-leakage currents of at least two orders of magnitude smaller. No clear field-effect was observed in the applied voltage range, in contrast to what it has been recently reported by Uekusa et al. \cite{Uekusa2020} with a series of TMB Acceptor CT crystals and films. Crystal bulk conductivity is probably dominant in the measured current, and the contribution of a possible thin-gate induced transistor channel at the dielectric/semiconductor interface is not clearly detectable. Estimated bulk conductivity values for sample 1 and sample 2 were $\sigma_{1}$=4$\cdot$10$^{-4}$ S/cm and  $\sigma_{2}$=2.2$\cdot$10$^{-4}$ S/cm, respectively.
The dc conductivity, measured at room $T$ at low current, is around $3 \cdotp 10^{-4}~ \Omega^{-1}$ \cm. The $T$ dependence of the conductance, evaluated as detailed in the Supporting Information, is reported for two samples in Figure \ref{fig:tmb-tqf4sigmadc}, and allows the extraction of the activation energy $E_a$, which turns out to be 0.32 -- 0.33 eV, in agreement with Ref. \cite{Uekusa2020}.

\begin{figure}[ht]
	\centering
	\includegraphics[width=0.8\linewidth]{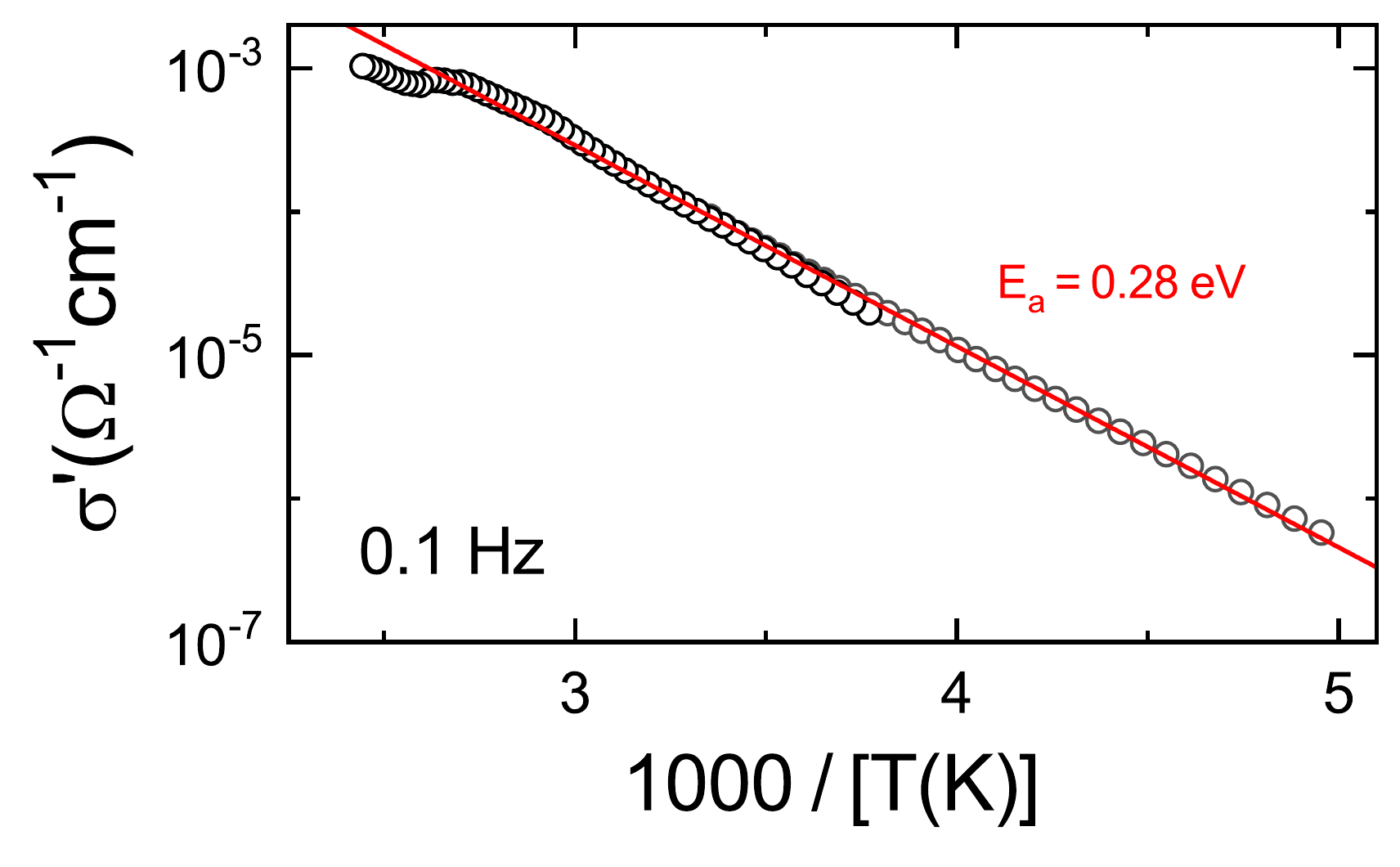}
	\caption{Temperature dependence of the TMB-\TF\ conductivity obtained
		by low-frequency dielectric measurements.}
	\label{fig:tmb-tqf4sigma}
\end{figure}

The temperature dependence of the conductivity can also be obtained
from low-frequency dielectric measurements, as reported in Fig. \ref{fig:tmb-tqf4sigma}. At such low-frequency
(0.1 Hz), $\sigma'(\omega,T)$ represents an estimate of the dc conductivity.
At 300 K the conductivity value is around $9 \times 10^{-5}~ \Omega^{-1}$ \cm, whereas the activation energy, estimated by applying the least square fitting
to the linear part of the conductivity shown in Fig. \ref{fig:tmb-tqf4sigma}, is 0.28 eV. 
In conclusion, both dc and low-frequency conductivity measurements
classify TMB-\TF\ crystal as a narrow-gap semiconductor ($E_a \sim 0.3$ eV), with a relatively high conductivity at room $T$ ($\sim 10^{-4}~ \Omega^{-1}$ \cm ).

\begin{figure}[h]
	\centering
	\includegraphics[width=0.9\linewidth]{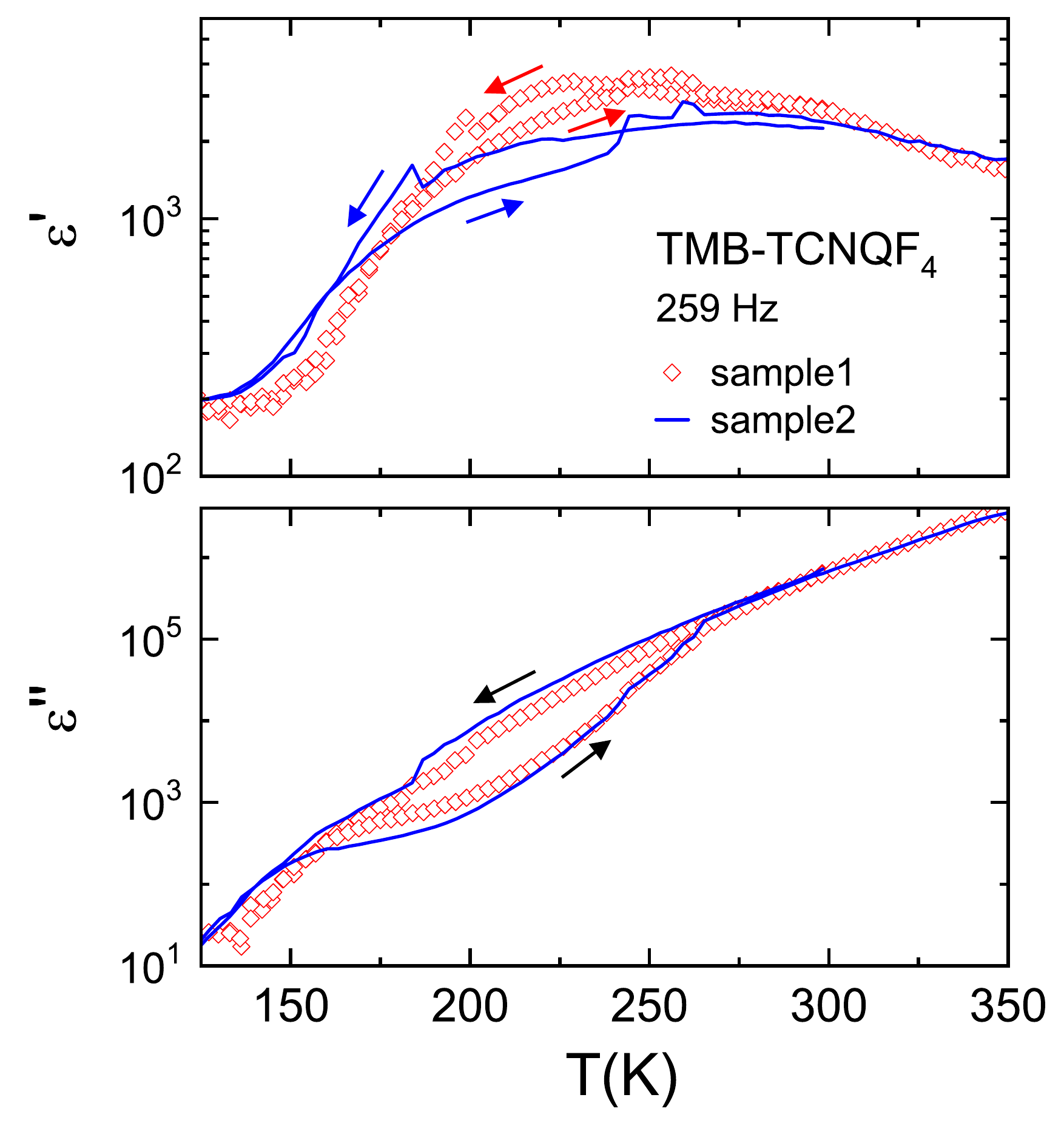}
	\caption{Temperature dependence of the real and imaginary part
		of the dielectric constant of two TMB-\TF\ samples at 259 Hz.
		The temperature cycle evidence a sort of hysteretic behavior
		between 270 and 170 K.} 
	\label{fig:tmb-tqf4hysteresis}
\end{figure}

Additionally, the dielectric measurements as a function of $T$ and frequency revealed a strange hysteretic behavior,
as exemplified by the $T$ dependence of the real ($\epsilon'$) and imaginary ($\epsilon''$) parts of the dielectric constant at 259 Hz shown in Fig. \ref{fig:tmb-tqf4hysteresis}. 
The minor anomalies observed in both quantities seem to look quite similar in two different samples, which is
puzzling and speaks against a non-intrinsic origin. We do not have any  evidence of a phase transition, and in any case the width of the hysteresis seems too large for it. 

\begin{figure}[ht]
	\centering
	\includegraphics[width=0.9\linewidth]{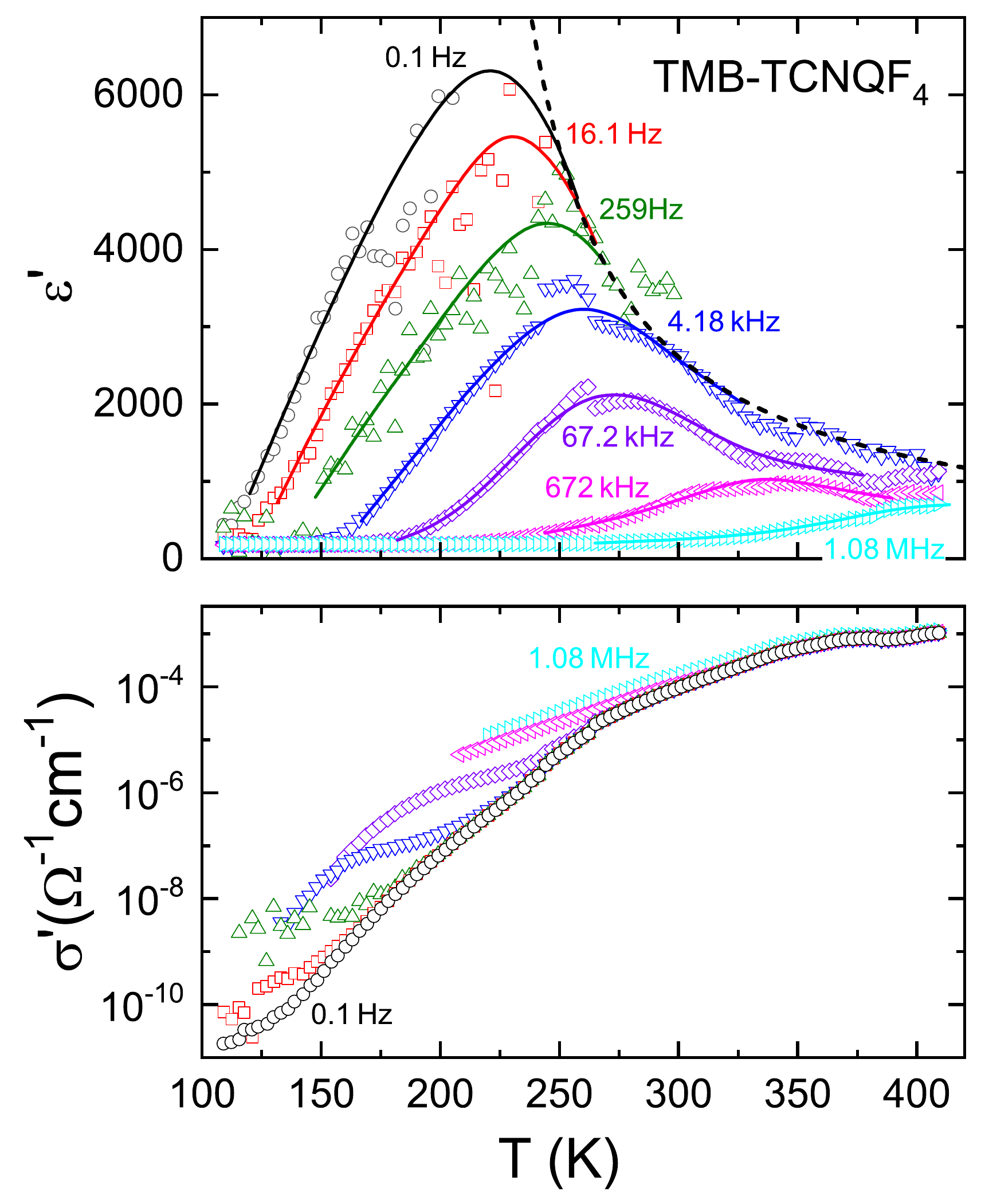}
	\caption{Top panel: temperature dependence of the real part of the dielectric constant $\epsilon'(T)$. The solid lines are guides for
		the eyes. The dashed line demonstrates Curie–Weiss behavior with $T_{CW} = 202$ K. Bottom panel: temperature dependence of the frequency dependent conductivity $\sigma'(T)$.}
	\label{fig:tmb-tqf4eps-sigma}
\end{figure}

In the top panel of Fig. \ref{fig:tmb-tqf4eps-sigma} the temperature-dependent
real part of the dielectric constant is shown for frequencies between 0.1 Hz and 1.08 MHz.
We observe large peaks in the permittivity, that decrease in amplitude and
shift to higher temperature with increasing frequency.
This corresponds to the typical behavior of relaxor ferroelectrics \cite{Cross1987,Samara2003}. Due to the needle-like geometry, the electrode
area and thus the measured capacitance are very small, leading to a large
uncertainty in the absolute values of $\epsilon'$.  Finally,  
the dashed line in the top panel of Fig. \ref{fig:tmb-tqf4eps-sigma} 
demonstrates that the right flanks of the relaxor peaks, representing the
static dielectric constant, can be described by a Curie-Weiss law with a Curie-Weiss temperature of $T_{\mathrm{CW}} \approx$ 202 K, which provides an estimate of the quasi-static freezing temperature. 
The $\sigma'(T)$ plots reported
in the bottom panel of Fig. \ref{fig:tmb-tqf4eps-sigma} reveal 
indications of the loss peaks, expected for relaxors. Their right flanks are partly superimposed by the dc conductivity, approximated by the 0.1 Hz curve (cf. Fig. \ref{fig:tmb-tqf4sigma}). 

\begin{figure}
	\centering
	\includegraphics[width=0.85\linewidth]{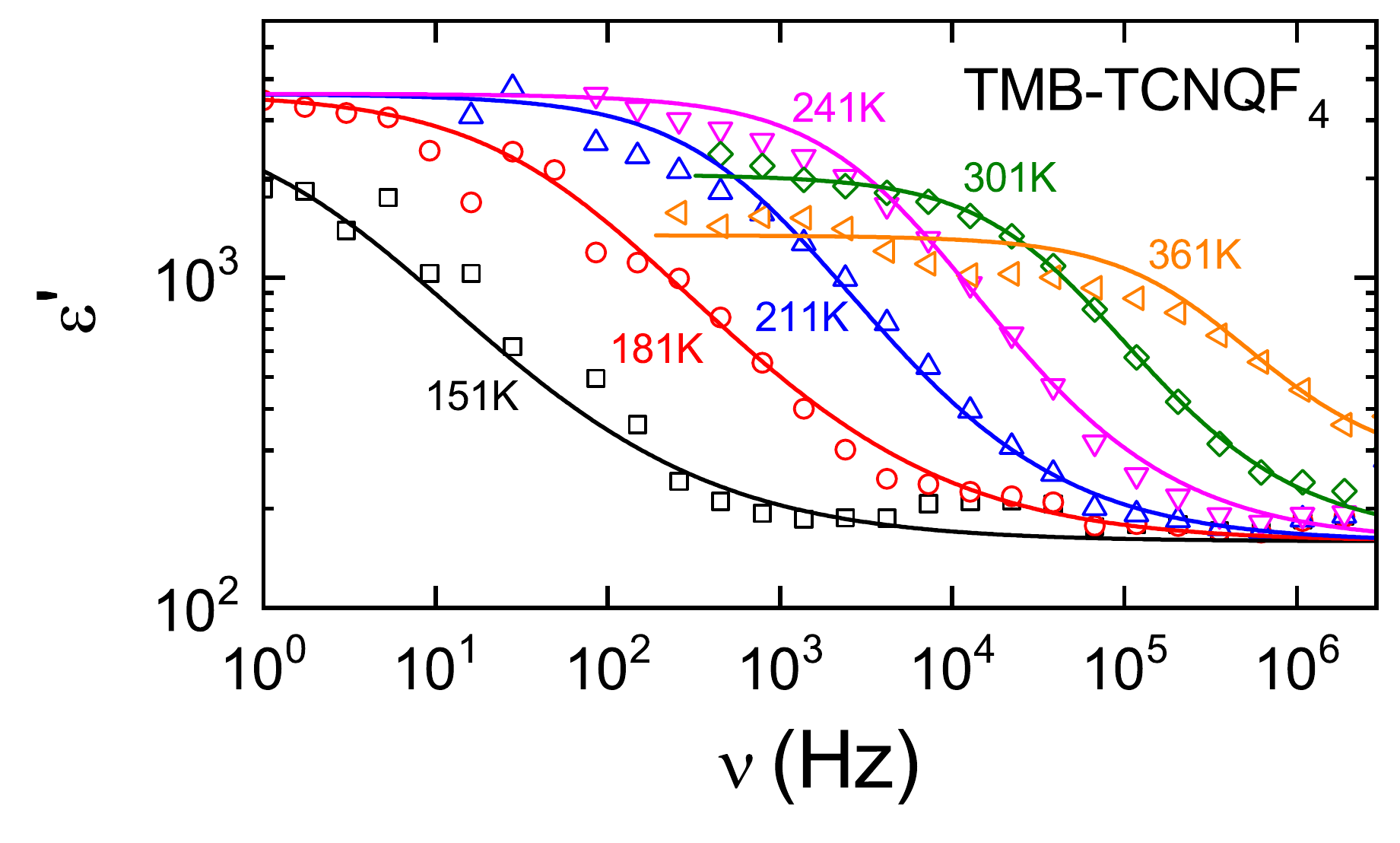}
	\caption{Frequency-dependent plot of the dielectric constant $\epsilon'(\nu)$ of TMB-\TF\ at various temperatures. Lines are fits with the  Cole-Cole function.}
	\label{fig:eps1nut}
\end{figure}

Frequency-dependent plots of the dielectric constant $\epsilon'(\nu)$ are
shown in Fig. \ref{fig:eps1nut} for various temperatures. The spectra reveal a step-like decrease of $\epsilon'(\nu)$ which shifts to lower frequencies with decreasing
temperature. This 
evidences the slowing down of relaxational
dynamics with decreasing $T$. Similar to the peaks in $\epsilon'(T)$,
the heights of the curves in $\epsilon'(\nu)$ decrease with increasing temperature,
typical of relaxor ferroelectrics \cite{Cross1987,Samara2003}.  

\begin{figure}[ht]
	\centering
	\includegraphics[width=0.7\linewidth]{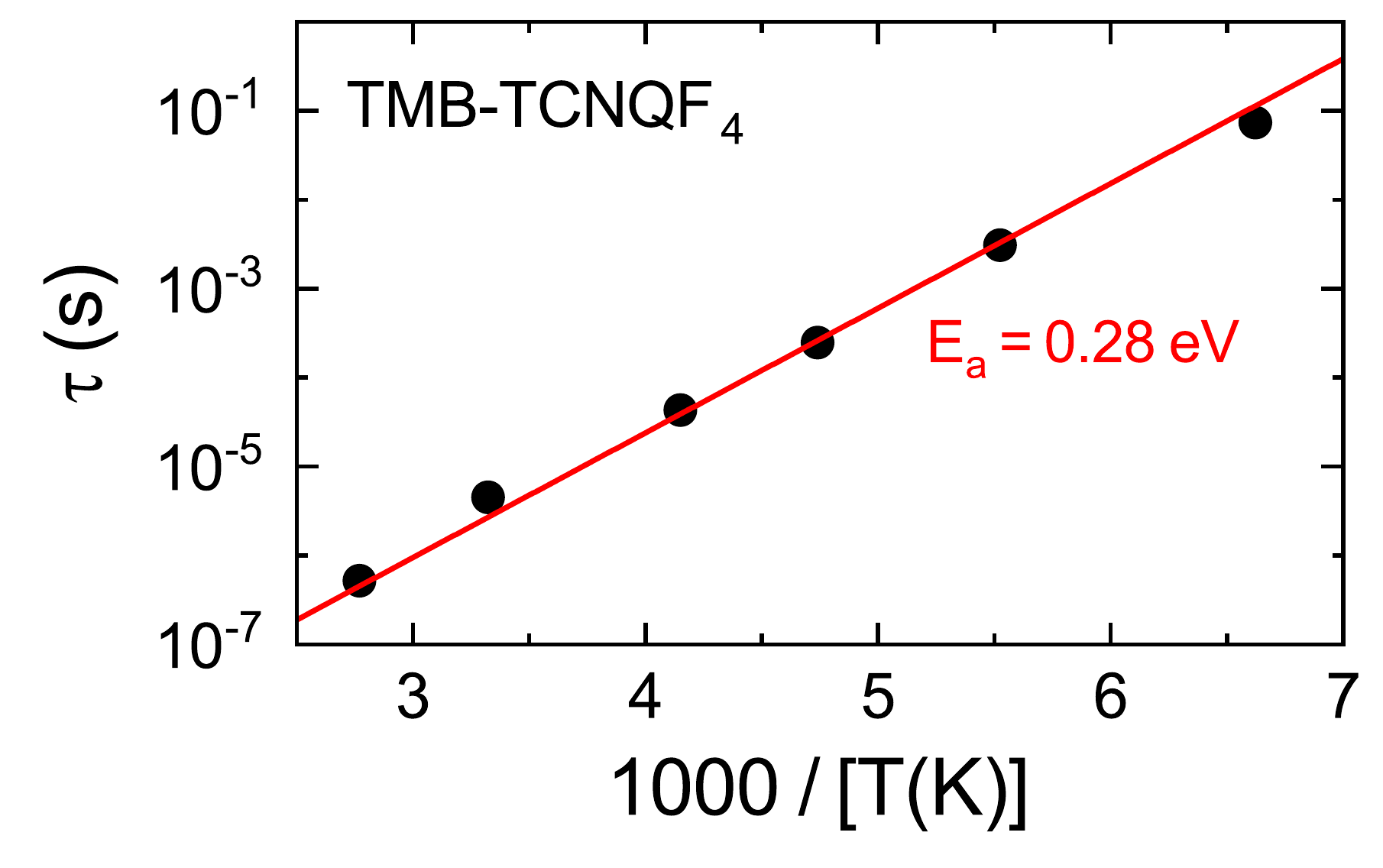}
	\caption{Temperature evolution of TMB-\TF\ relaxation time in an Arrhenius plot. The line shows an Arrhenius fit.}
	\label{fig:tmb-tqf4tauinveret}
\end{figure}

To further analyze the relaxor dynamics,
Fig. \ref{fig:tmb-tqf4tauinveret} presents an Arrhenius plot of the temperature-dependent relaxation times as determined from the fits shown in Fig. \ref{fig:eps1nut}.
The linear Arrhenius fit (line in the figure) yields an
activation energy $E_a \sim$ 0.28 eV, and a pre-exponential factor
$\tau_0 =5.9 \times 10^{-11}$ s. 
Similar Arrhenius behavior we have also observed for another organic relaxor, M$_2$P-TCNQ \cite{Fischer2021}, whereas most, but not all, relaxor ferroelectrics can be described by the Vogel-Fulcher-Tammann law \cite{Vogel1921,Fulcher1925,Tammann1900}. It is interesting to note that we have
determined identical energy barriers for the dipole motion and the dc charge
transport, indicating a close coupling of both dynamics.  

Before closing this Section, a word of caution is necessary. In fact,
one must be aware that a non-intrinsic, contact-related origin of the found relaxation
process cannot be completely excluded. The usual tests to check for
this effect, namely measurements with different sample geometries and with very
different contact types cannot be easily done for the present needle-like, very brittle samples. Due to the sample geometry
and the conductivity of the samples, we also cannot apply very high fields
to check for hysteresis or polarization in the presumed relaxor ferroelectric
state. However, the presence of (partially) ordered dipoles implied in a
ferrolectric behavior is consistent with the lack of inversion center in the
stack, as suggested by the reported vibrational measurements.\cite{Castagnetti2018}

\subsection{Temperature dependence of the vibrational spectra and revision of the crystal structure}
According to the reported crystal structure \cite{Castagnetti2018},
at room temperature TMB-\TF\ crystallizes in the monoclinic system
$C2/m$ ($C_{2h}^3$), with two DA pairs per unit cell.
All the molecules reside on inversion
centers so that the stack appears to be regular. On the other 
hand, it has been also noted that the room temperature IR spectra polarized parallel to the stack are characterized by the presence of strong IR absorptions induced by the electron-molecular vibration (e-mv) coupling at the same frequencies of the main Raman bands \cite{Castagnetti2018}. This is an unquestionable hint of the loss
of inversion center, i.e., the stack appears to be dimerized, in contrast to the X-ray crystal structure.

\begin{figure}[h]
	\centering
	\includegraphics[width=0.85\linewidth]{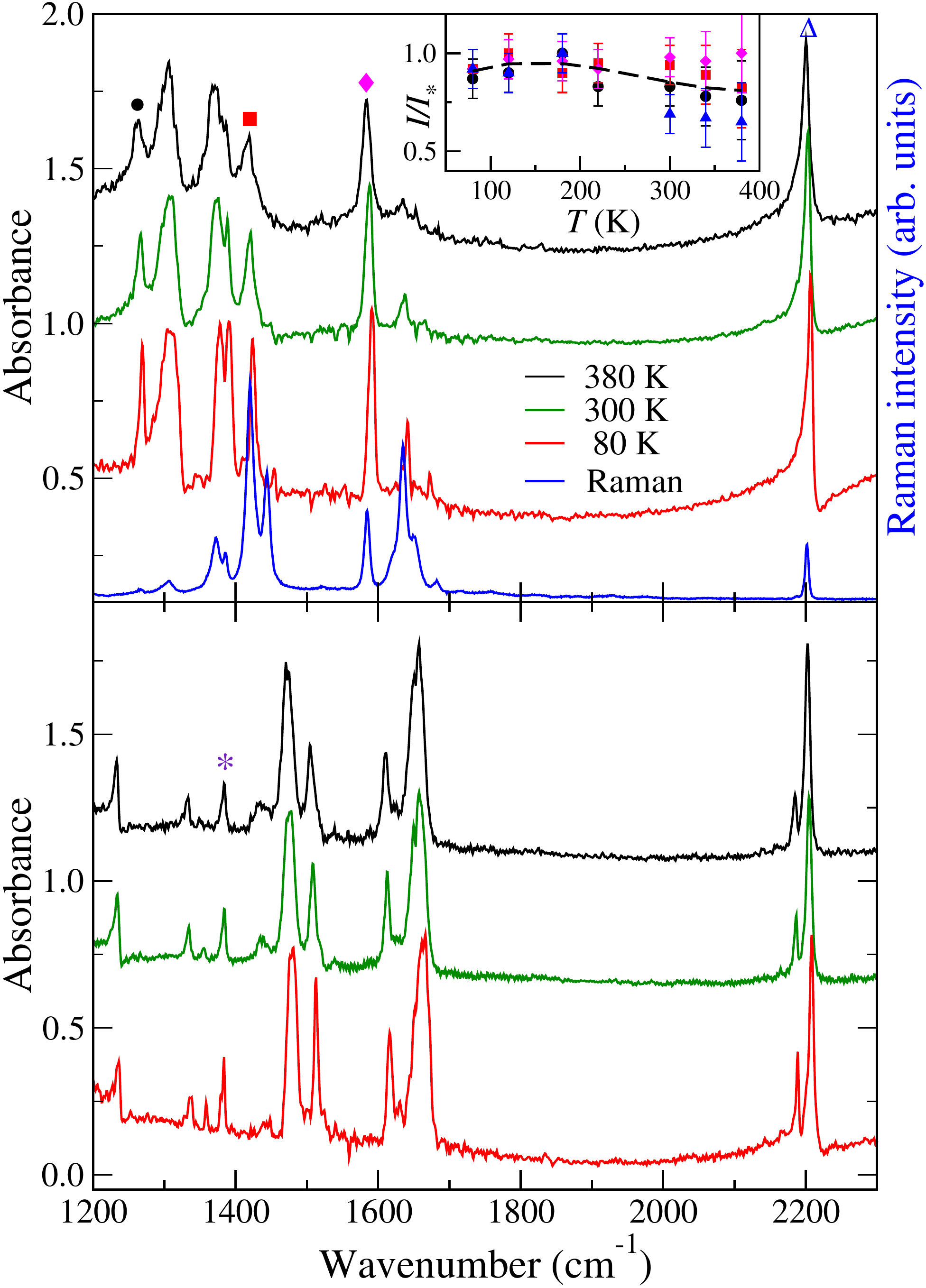}
	\caption{Infrared spectra of TMB-\TF\ as a function of $T$. 
		Top panel: Spectra polarized parallel to the stack axis,
		compared with the 300 K Raman one (blue) to evidence the vibronic bands.
		Bottom panel: Spectra polarized perpendicular to the stack axis. Inset of top panel: $T$ evolution of the ratio of the intensity of four e-mv induced bands (marked as black dot, red square, magenta diamond and blue triangle) with respect to a normally infrared active band (marked by an asterisk in the bottom panel).}
	\label{fig:tmb-tqf4_ir_tdep}
\end{figure}

There are several other cases of inconsistency between the results of
X-ray and of vibrational spectroscopy, since the former probes long-range
order and the latter the local (DA pair) structure. All these
discrepancies have been explained in terms of some kind of disorder,
either static or dynamic \cite{Girlando1984,Girlando1988,Bewick2005,Kumar2011}.
In the case of TMB-\TF, lack of inversion center and disorder at the nanoscale
are also implied by the relaxor behavior illustrated
in the previous Section. We have then decided to examine the temperature dependence
of the vibrational spectra, to investigate if a disorder-to-order phase transition
was present, like in the case of two other examples we are aware of \cite{Girlando1984,Bewick2005}.

With IR spectroscopy, we could go both above and below room temperature,
as exemplified in Figure \ref{fig:tmb-tqf4_ir_tdep}. No visually appreciable
band intensity variations are detected between 380
and 80 K. To confirm this finding, we measured the relative intensity of four e-mv induced bands (marked in the top panel of the Figure as a black circle, a red square, a magenta diamond and a blue triangle) with respect to a normally IR active band detected in the spectrum polarized perpendicularly to the stack (marked by an asterisk in the bottom panel of the Figure). The results are plotted in the inset of Figure \ref{fig:tmb-tqf4_ir_tdep}. The average relative intensity shows a slight increase in going below about 
220 K, but the change is well within the estimated error bars, which are larger at high $T$ since the bands become broader. Therefore there is no clear-cut
evidence of a disorder-to-order transition by lowering $T$, but rather a behavior similar
to that observed for the temperature dependence of the dielectric constant (Figure \ref{fig:tmb-tqf4hysteresis}).
\begin{figure}
	\centering
	\includegraphics[width=0.85\linewidth]{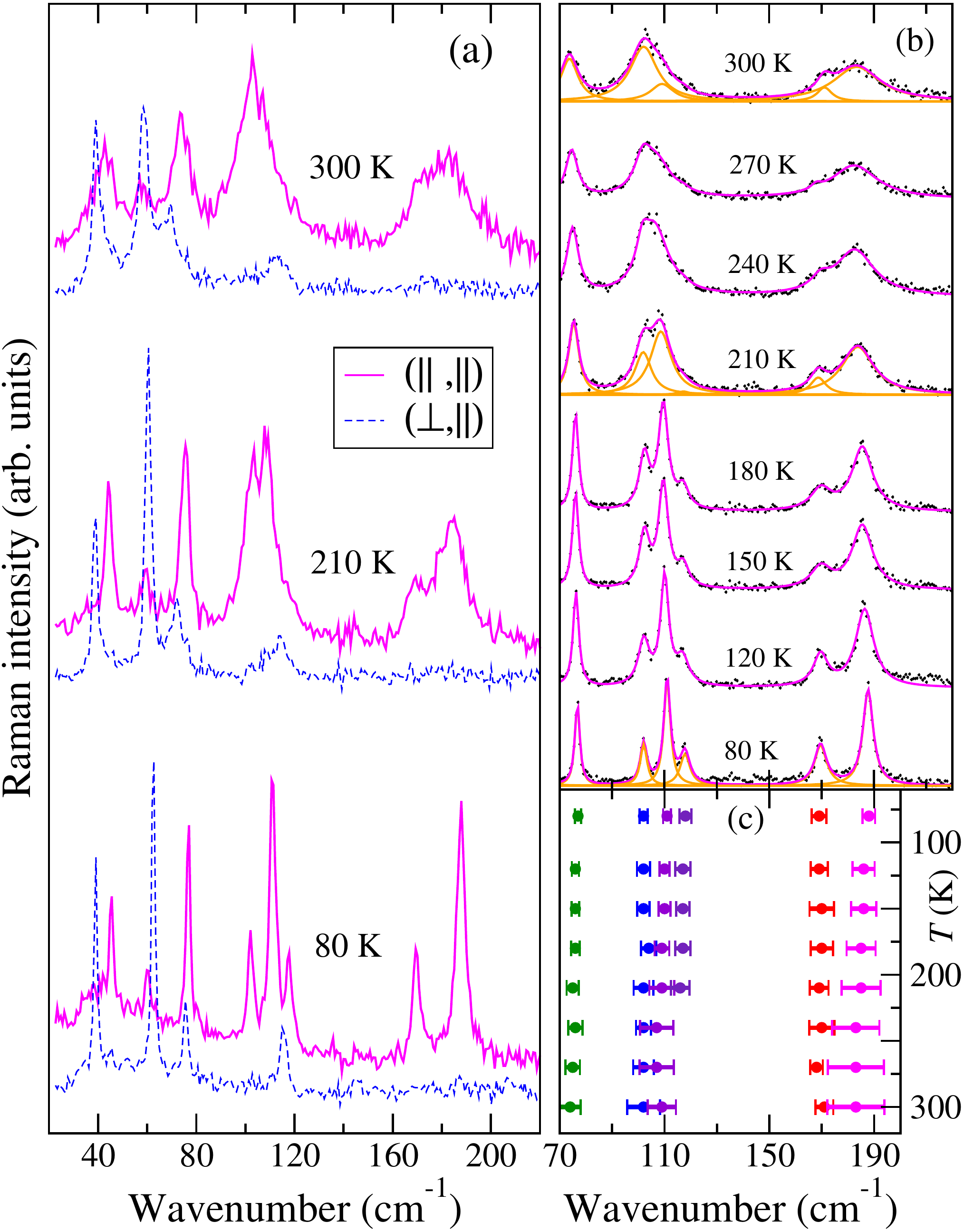}
	\caption{Low-frequency Raman spectra as a function of temperature.
		(a) Spectra at three significant temperatures. The $\parallel$ and $\perp$ symbols within parenthesis indicate the polarization of incident and scattered light with respect to the stack axis. (b) Full $T$ evolution
		of an enlarged portion of the ($\parallel,\parallel$) spectra. Black
		dots represent the experimental spectra, magenta lines the fitting
		with a set of Lorentzian functions. The single Lorentzians are reported as orange lines for the three most significant temperatures.
		(c) $T$ evolution of the peak frequencies (dots) and of the 
		corresponding FWHM (full width at half maximum, here represented as error bars)	of the bands of panel (b).}
	\label{fig:RamanT_alli}
\end{figure}

Raman spectroscopy in the low-frequency (10--200 \cm) region, involving 
lattice phonons, is known to be very sensitive to molecular
packing \cite{Brillante2008} and also to disorder in a scale of a few
unit cells \cite{Brillante2012}, hence intermediate between the 
long-range order of X-ray and the local structure probed by high-frequency, intramolecular vibrations. We have thus obtained the low-frequency polarized Raman spectra of TMB-\TF\ as a function of temperature shown in Figure \ref{fig:RamanT_alli}.

The selection rules for TMB-\TF\ Raman active lattice phonons in terms of the
$C2/m$ ($C_{2h}^3$) factor group and adopting the rigid molecule approximations \cite{Turrell1972}, are as follows: 2 $A_g$ + 4 $B_g$. So we expect
only six bands, corresponding to the librations of the
molecules, since the primitive cell contains just one DA pair,
with each molecule on inversion center. A look at the experimental
spectra shows that these spectral predictions are not obeyed, as there
are more bands than expected. The highest frequencies bands, above 160 \cm, might also involve the methyl rotations, but even without counting them 
we observe eight bands, which become 10 at 80 K (Panel (a) of Fig. \ref{fig:RamanT_alli}). Therefore the lattice mode spectral region,
like that of the intramolecular vibrations, 
tells us that the previously reported space group \cite{Castagnetti2018} is not the correct one.

In addition, at room temperature the spectra recorded
with incident and scattered light polarized along the stack, labeled
($\parallel,\parallel$), show two rather broad bands around 100 and
180 \cm, which by lowering $T$ narrow and separate in more components,
as detailed in panel (b) of the Figure. Notice also that the
bandwidths observed in the other polarization are normal.
This situation is reminiscent of what it has been observed in
another CT crystal, \cite{Fischer2021}. In that case,
the band broadening has been attributed to the electron-phonon coupling, since the broad bands also exhibit a marked frequency hardening
by increasing the temperature, and their intensity is strongly
enhanced by moving the frequency of the Raman exciting line towards
the far-red, going into pre-resonance with the CT transition.
Strong electron-phonon coupling means strong phonon
anharmonicity, hence the line broadening.   

To test if this explanation is valid also for TMB-\TF, 
we have followed in detail the temperature evolution of
Raman bands in the ($\parallel,\parallel$) polarization, 
by performing a deconvolution in terms of Lorentzians, as exemplified for three 
temperatures by the orange lines in panel (b). Panel (c) summarizes
such an analysis, in terms of the peak frequencies and Lorentzian FHMV, the latter represented by error bars. It is seen that the
above-mentioned two groups of bands around 100 and 180 \cm~
start to separate around 200 K, but no anomalous
frequency hardening is evident by lowering the temperature.  
Furthermore, the spectra recorded by moving the exciting
line toward the CT transition (see Supporting Information, Figures S1 and S2)
show an intensity enhancement
much less pronounced than in the case of M$_2$P-TCNQ \cite{Fischer2021}.
Therefore in this case the remarkable band broadening
and line merging observed by going towards room temperature
cannot be fully ascribed to the effect of electron-phonon coupling. 
In other words, whereas the anomalous band broadening
observed in M$_2$P-TCNQ is mostly due to phonon anharmonicity,
and can be assimilated to the so-called dynamic or thermal
disorder, the analogous broadening we observe in TMB-TCNQF$_4$
must have a different origin. We believe it is due
to disorder in the lattice structure along the stack axis,
namely, to static, or displacement, disorder \cite{guinier_book}.

In summary, low-frequency Raman spectroscopy, like IR, tells us that the space group symmetry is \textit{lower} than that obtained in the standard refinement of
the room temperature X-ray data \cite{Castagnetti2018},
and that above  $\approx$ 200 K there is some kind of disorder along the stack.

In order to clarify the apparent discrepancy of X-ray with respect to vibrational and dielectric
measurements, we used synchrotron radiation
for a new structural analysis,
carefully inspecting the reciprocal space of the samples to look for signs
of diffuse scattering, of phase transitions, or of altered symmetry.
This set of measurements
was limited to cryogenic conditions (100, 150 and 200 K), to avoid the risk of beam damage.

Overall, no significant diffuse scattering features were found, which could agree with correlated disorder along the molecular stacking direction, while a strong crystal mosaicity elongates all reflections, hampering a reliable diffuse scattering analysis. However, the presence of strong reflections violating a $C$ centering of the unit cell (cf. Figure S7) proved that the symmetry is indeed lower than the previously reported one \cite{Castagnetti2018}. Successful indexing of all reflections was achieved with space group $P2_1/m$ rather than $C2/m$ \cite{Castagnetti2018}, while maintaining the same unit cell metric and orientation. Full details of the new crystallographic analysis can be found in the Supporting Information.

\begin{figure}[hp]
	\centering
	\includegraphics[width=\linewidth]{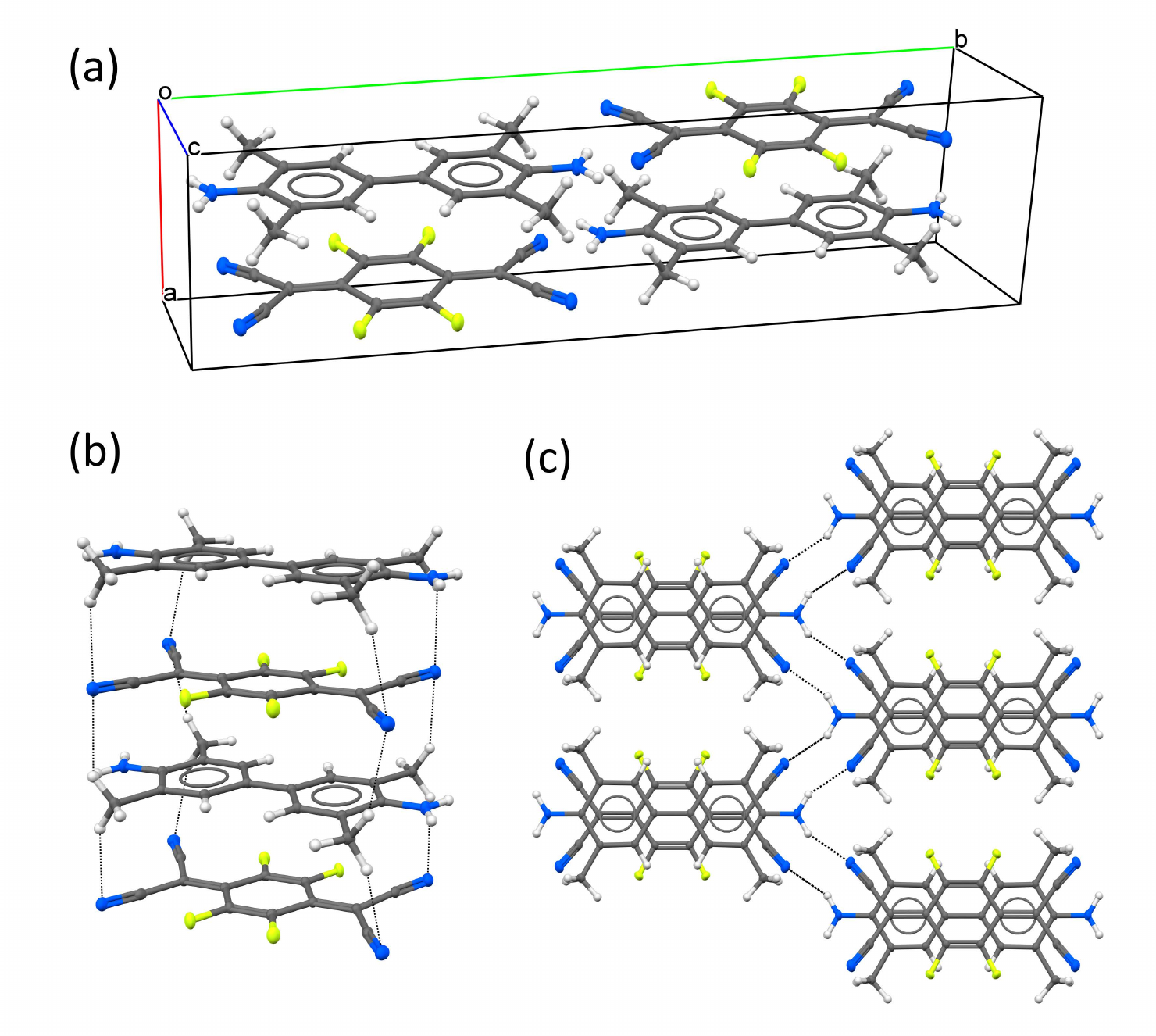}
	\caption{Structure of TMB-TCNQF$_4$ at 100 K. The main intra- and inter-stack
		hydrogen bonding  between TMB and TCNQF$_4$ are shown in (b) and (c), respectively.}
	\label{fig:tmb-tcnqf4struct100k}
\end{figure}

The $P2_1/m$ structure features two stacks (two formula units) per unit cell, aligned along the $a$ axis (Figure \ref{fig:tmb-tcnqf4struct100k}). Most importantly, and differently from what was modeled by using $C2/m$ space group, these stacks are dimerized, and the inversion center
is \textit{between} the stacks. Therefore the two DA dimers in the unit cell have  
anti-ferroelectric arrangement.
The intra-dimer and inter-dimer distances along the stack at 100 K are 3.1447(10)
and 3.4415(11), so that the dimerization amplitude $\delta = (d_2-d_1)/(d_2+d_1)$ is
0.05 (Figure S5), about twice that of the of the prototypical CT crystal Tetrathiafuvalene-Chloranil (TTF-CA) in the low temperature phase \cite{Cointe1995}. 
The crystal structure also evidences the presence of a network of hydrogen bonds between the methyl and the cyanide groups (Figure \ref{fig:tmb-tcnqf4struct100k} (b)). These cooperate with the CT stacking interactions, while more directional H-bonds contacts are found between cyanide and amino groups of molecules belonging to different stacking columns (Figure \ref{fig:tmb-tcnqf4struct100k} (c)). All these intermolecular interactions can also be visualized in terms of the Hirshfeld surfaces provided by the program Crystal Explorer 17 \cite{Spackman2021}, as detailed in the Supporting Information.

The number of bands observed in the low-frequency Raman spectra (Figure \ref{fig:RamanT_alli})
is consistent with the revised $P2_1/m$ structure, as are the e-mv induced bands observed
in IR (Figure \ref{fig:tmb-tqf4_ir_tdep}), since the primitive cell contains two dimerized stacks.
We have collected a new set of X-ray data at 300 K, verifying that structure remains the same also at room temperature. Indeed, the weak extra peaks due to the primitive $P2_1/m$
lattice are also present in the room temperature X-ray diffraction patterns (Figure S7), and these were ignored in the previous automatic structure refinement.\cite{Castagnetti2018} 
Therefore there are no phase transitions up to 300 K. 
However, the new temperature dependent structural analysis suggests that something is
happening between 300 and 200 K. For instance, the dimerization amplitude $\delta$
increases from 0.04 to 0.05 by lowering $T$ in this interval, then it remains practically
constant down to 100 K (Figure S5), a behavior consistent with the increase of the
intensity of the e-mv induced bands (inset of Figure \ref{fig:tmb-tqf4_ir_tdep}).
Relaxor behavior and the low-frequency Raman spectra suggest the presence of
some kind of disorder above  $\approx$ 200 K.

We attempted an analysis of the temperature
dependence of the anisotropic displacement parameters, ADP \cite{Burgi2000}.
The ADP $T$ dependence gives some hint of disorder along the stack above 200 K,
but the fact that we do not have structural data between 200 and 300 K, and that the
300 K data have been collected with a different technique and on a different sample
with respect to the others, 
prevent us to conclude anything definitive. 
Indeed, the investigation of the disorder by structural methods, for instance by
pair-distribution function analysis, would require temperature dependent synchrotron radiation or neutron diffraction (in view of the above evidenced importance of hydrogen bonding) on a unique crystal having optimal crystallinity, and from ambient conditions to 100 K in finer steps. Such deep investigation is certainly desirable, as it would allow a direct and hopefully precise
connection with vibrational and dielectric measurements, but goes well beyond the
scope of the present work.

\section{Discussion and conclusions}

The measurements presented here show that the physics
of TMB-TCNQF$_4$ is quite interesting and peculiar, but also rather complex, so that
its understanding still requires additional experiments aimed to provide suitable data for a better modeling.
It has been shown that TMB-TCNQF$_4$ is a narrow gap semiconductor ($E_a \simeq 0.3$ eV),
with an optical gap of about 0.8 eV, and a quasi ionic ground state ($\rho \simeq 0.9$) \cite{Castagnetti2018}. 
The most important advancements of the present study with respect to the previous ones \cite{Castagnetti2018,Uekusa2020} consist in the finding that TMB-TCNQF$_4$ stacks
are dimerized, and that some kind of disorder along the stacks is likely present above $\approx$ 200 K,
a fact that is also reflected by the observed relaxor behavior. As it is generally
the case with relaxors,\cite{Ahn2016} the passage from the disordered to the ordered structure takes
a large $T$ interval, $\approx 100$ K  (Figs. \ref{fig:tmb-tqf4hysteresis}, \ref{fig:tmb-tqf4_ir_tdep}, and \ref{fig:RamanT_alli}), and does not look
like a real phase transition. Moreover, we have found that the room $T$ conductivity is relatively high ($\sim 10^{-4}~ \Omega^{-1}$ \cm),
and decreases by three orders of magnitude in going to 200 K. Finally, relaxor-ferroelectric-like behavior
has been observed already at room temperature, with a maximum in $\epsilon^\prime(\nu,T)$ around 200 K and 0.1 Hz 
(top panel of Fig. \ref{fig:tmb-tqf4eps-sigma}). The relaxor behavior is consistent with the disorder,
but not with the anti-ferrolectric arrangements of the two stacks in the unit cell.

Modeling of TMB-TCNQF$_4$ physical properties in terms of a Peierls-Hubbard model and/or band
structure relevant to a regular, ordered stack is clearly inappropriate \cite{Castagnetti2018,Uekusa2020}.
Electrical properties have to take into account the gap opening due to dimerization and the effect of disorder. Simple band structure calculations (Supporting Information) for the
100 K ordered structure show that the gap is due to dimerization and not to electron-electron interactions, so TMB-\TF~ has to be considered a \textit{band}, not Mott, semiconductor.
Further, deeper analyses are needed to understand the role of disorder above 200 K.

However, at the present stage we can envision
a quite plausible scenario able to reconcile all the presently available experimental data,
giving directions for future investigations.
Analysis of the low $T$ structures of TMB-TCNQF$_4$ has shown that the
system is strongly 1D, with the TMB and TCNQF$_4$ molecules perfectly aligned
on top of each other along the stacks, and weak inter-stack interactions.
In going above $\approx 200$ K, the structure becomes disordered {\it along}
the stack, as indicated by the low-frequency Raman spectra and consistent with the relaxor
behavior. This kind of disorder
is most likely displacitive, and adds to the usual thermal disorder.

\begin{figure}
	\centering
	\includegraphics[width=\linewidth]{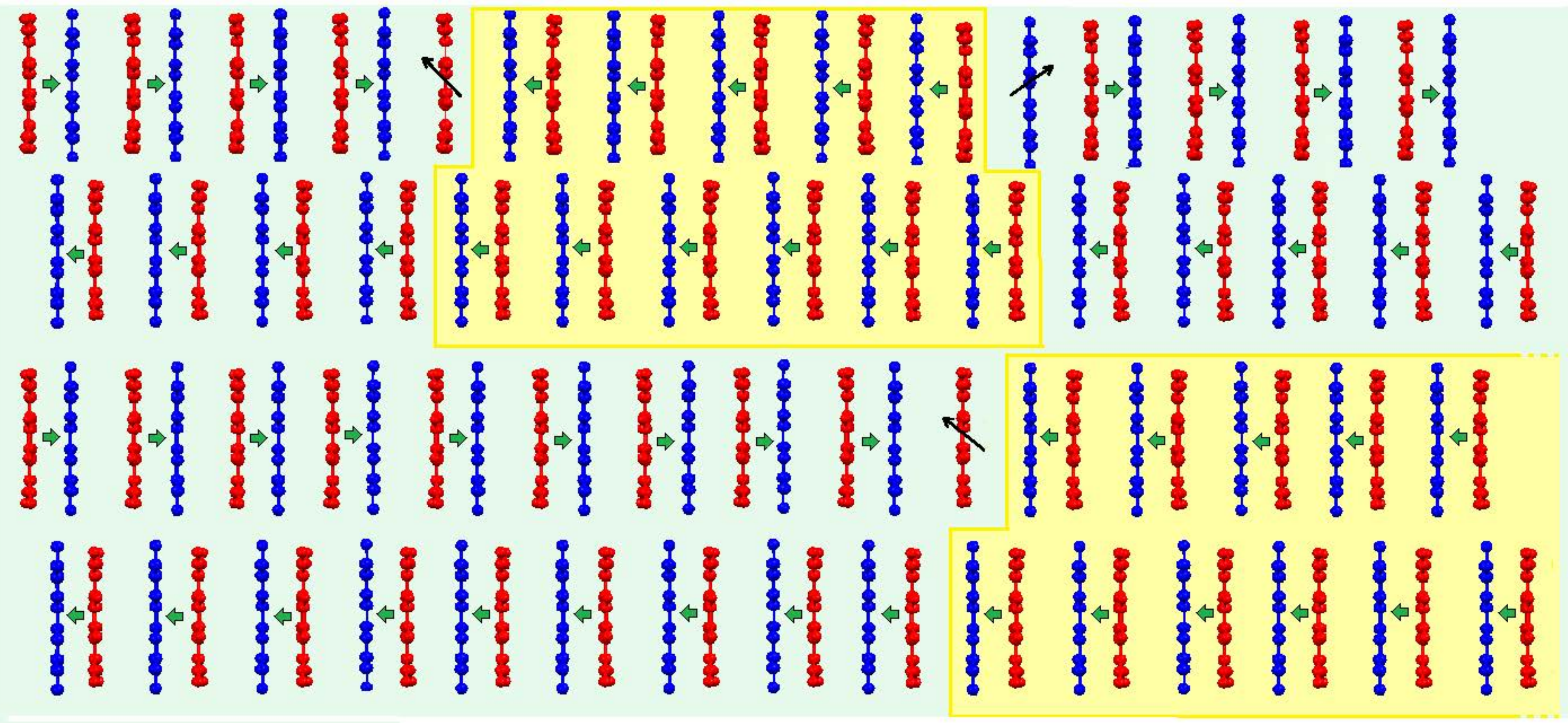}
	\caption{Illustration of how the presence of spin solitons
		in TMB-TCNQF$_4$ yields ferroelectric domains (pale yellow regions) inside a
		otherwise anti-ferrolectric arrangement (light green regions). Blue molecules: TMB;
		red molecules: TCNQF$_4$. The dimerization is amplified for clarity. The green bold arrows
		indicate the direction of the electric dipole within the dimer, whereas the
		spin soliton is indicated by an inclined black arrow on the molecule.}
	\label{fig:spinsoliton}
\end{figure}

As we stated in the Introduction, 1D
systems like the present one are subject to the spin-Peierls instability,
which is the driving force of the dimerization. However, it is also well known 
that in strictly 1D systems
phase transitions are not allowed down to 0 K, and that fluctuations are present
until 3D interactions become significant \cite{Pincus1975}. 
In TMB-TCNQF$_4$ the dimerization is probably also affected by the methyl-nitrogen contacts,
(see Figure \ref{fig:tmb-tcnqf4struct100k}) but here we are proposing a rather simplified general scenario.
Fluctuations involve ``defects''  of dimerization, boundaries between
domains of different orientation, namely the spin solitons depicted in
Fig. \ref{fig:spinsoliton}. At room $T$ the disorder is likely due to
presence of nanoscopic domains, which are able to move under the effect of the electric
field. And there is no correlation between domains in different chains,
so that we have the possibility of ferroelectric regions (indicated in pale yellow in the Figure) inside an anti-ferroelectric structure. In this model, these ferroelectric nanodomains 
are the origin of the relaxor behavior, in accord with the common explanation of relaxor ferroelectricity by nanoscale ferroelectric order \cite{Cross1987,Samara2003}.
By lowering $T$, the population of solitons decreases,
namely, the dimension of ferroelectric domains increases and their motions under
the electric field becomes slower and slower until the electrostatic interactions
and inter-chain hydrogen-nitrogen contacts lock the chains into the fully ordered
anti-ferroelectric  $P2_1/m$ structure.

Besides polyacetylene \cite{Heeger1988}, spin solitons have been detected and
widely studied in the CT crystal TTF-CA and derivatives in their low temperature or high pressure ($p$) dimerized stack phase \cite{Kagawa2010,Sunami2018,Soos2007}. Although the present data do not fully prove it, the above scenario is then quite 
plausible, suggesting a wide playground
of new experiments and theory, perhaps more promising than that of TTF-CA.
In fact, whereas in TTF-CA the studies have to be performed at low $T$ or high
$p$, here solitons are possibly present at ambient conditions. Since
TMB-TCMQF$_4$ is quasi-ionic, triplet spin excitations are likely thermally
populated and easily experimentally accessible. Proper characterization of the disorder, for instance by neutron diffraction,
would allow establishing the correlation length, or dimensions of the differently
oriented nanodomains, which would provide the basis for a better understanding
of the still rather mysterious relaxor ferroelectric behavior.
Efforts should be also devoted to the growth of the
crystals by different methods, possibly in the form of thin films. If successful,
TMB-TCNQF$_4$ might be interesting even from the perspective of materials application.
In fact, the dimerized phase of TTF-CA has a ferroelectric arrangement, and poling
yields to an ordered phase with ferroelectric hysteresis \cite{Kobayashi2012}.
In TMB-TCNQF$_4$ the ordered state is anti-ferroelectric, but
the ferroelectric domains would offer the possibility of building a room $T$
relaxor ferroelectric \cite{Ahn2016}.

\section*{Acknowledgments}

AG thanks Prof. Pascale Foury-Leylekian for very helpful discussions about the crystallographic issues. RF thanks Prof. Anthony Linden for his help in the X-ray diffraction data collection.
JKHF and PL acknowledge funding from the Deutsche Forschungsgemeinschaft (DFG) via the Transregional Collaborative Research Center TRR80 (Augsburg, Munich). R.P. and M.M-T. acknowledge support from the Marie Curie Cofund, Beatriu de Pin\'os Fellowships (Grant Nos. AGAUR 2017 BP 00064). This work was also supported by the Spanish Ministry project GENESIS PID2019-111682RB-I00, the “Severo Ochoa” Programme for Centers of Excellence in R\&D (FUNFUTURE, CEX2019-000917-S) and the Generalitat de Catalunya (2017-SGR-918). The Elettra Synchrotron (CNR Trieste) is acknowledged for granting the beamtime at the single-crystal diffraction beamline XRD1 (Proposal ID 20185483). In Parma the work has benefited from the equipment and support of the COMP-HUB Initiative, funded by the ``Departments of Excellence'' program of the Italian Ministry for Education, University and Research (MIUR, 2018-2022).

\bibliography{TMB-TCNQF4}

\newpage
\pagestyle{empty}

\begin{figure}[htp] \centering{
\includegraphics[scale=0.95]{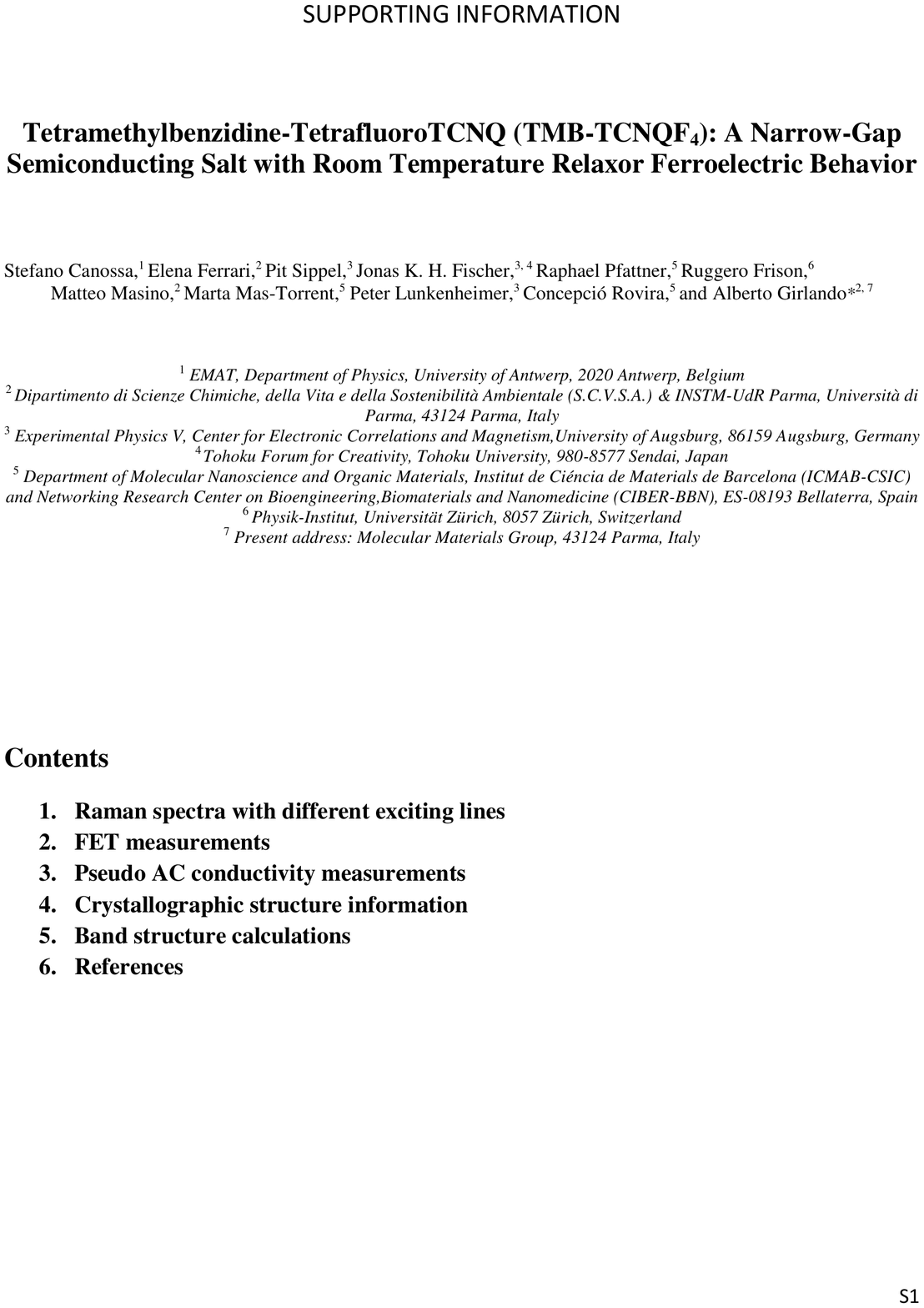}}
\end{figure}
\begin{figure}[htp]
	 \centering{
\includegraphics[scale=0.95]{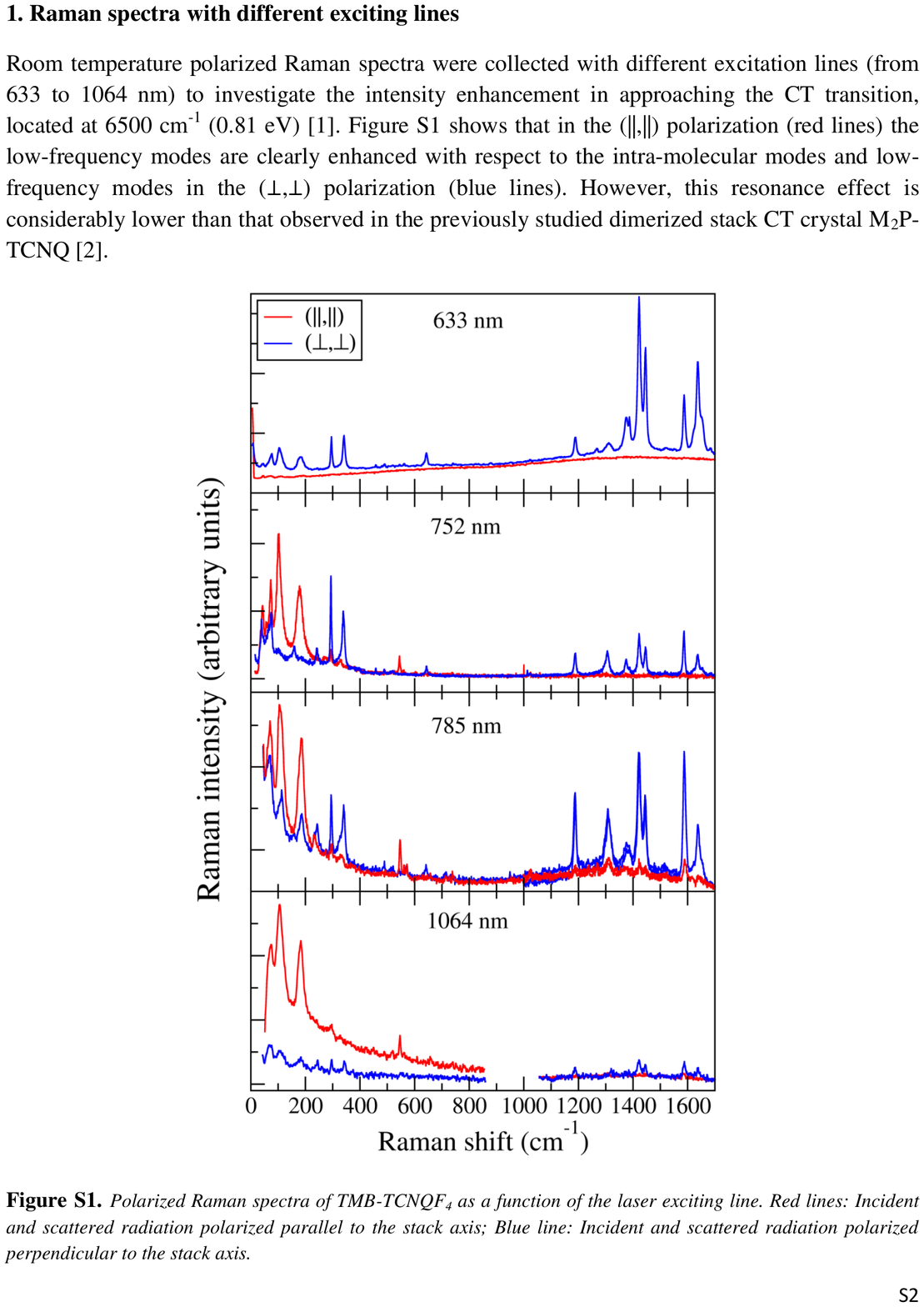}}
\end{figure}
\begin{figure}[htp]
	\centering{
		\includegraphics[scale=0.95]{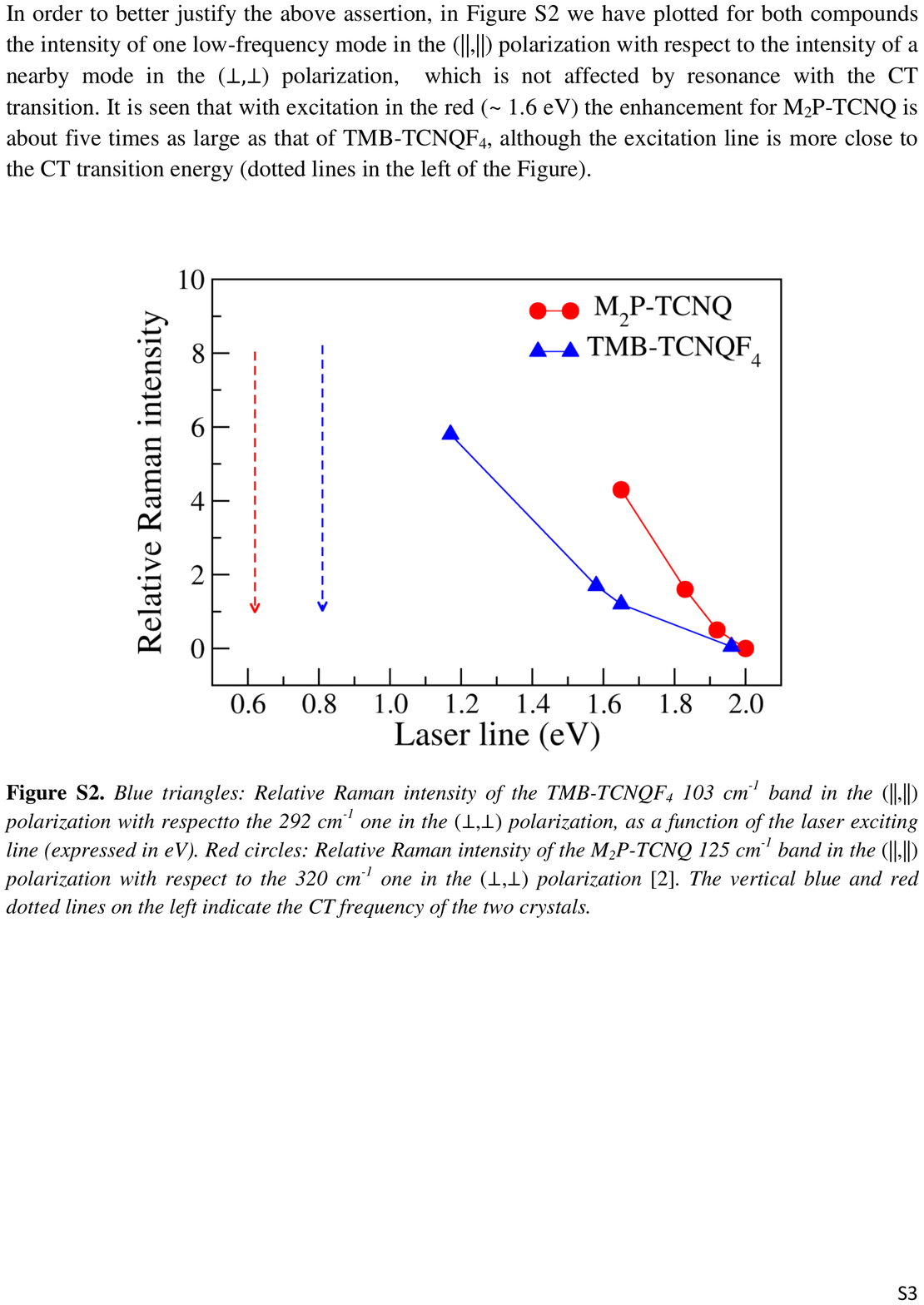}}
\end{figure}
\begin{figure}[htp]
	\centering{
		\includegraphics[scale=0.95]{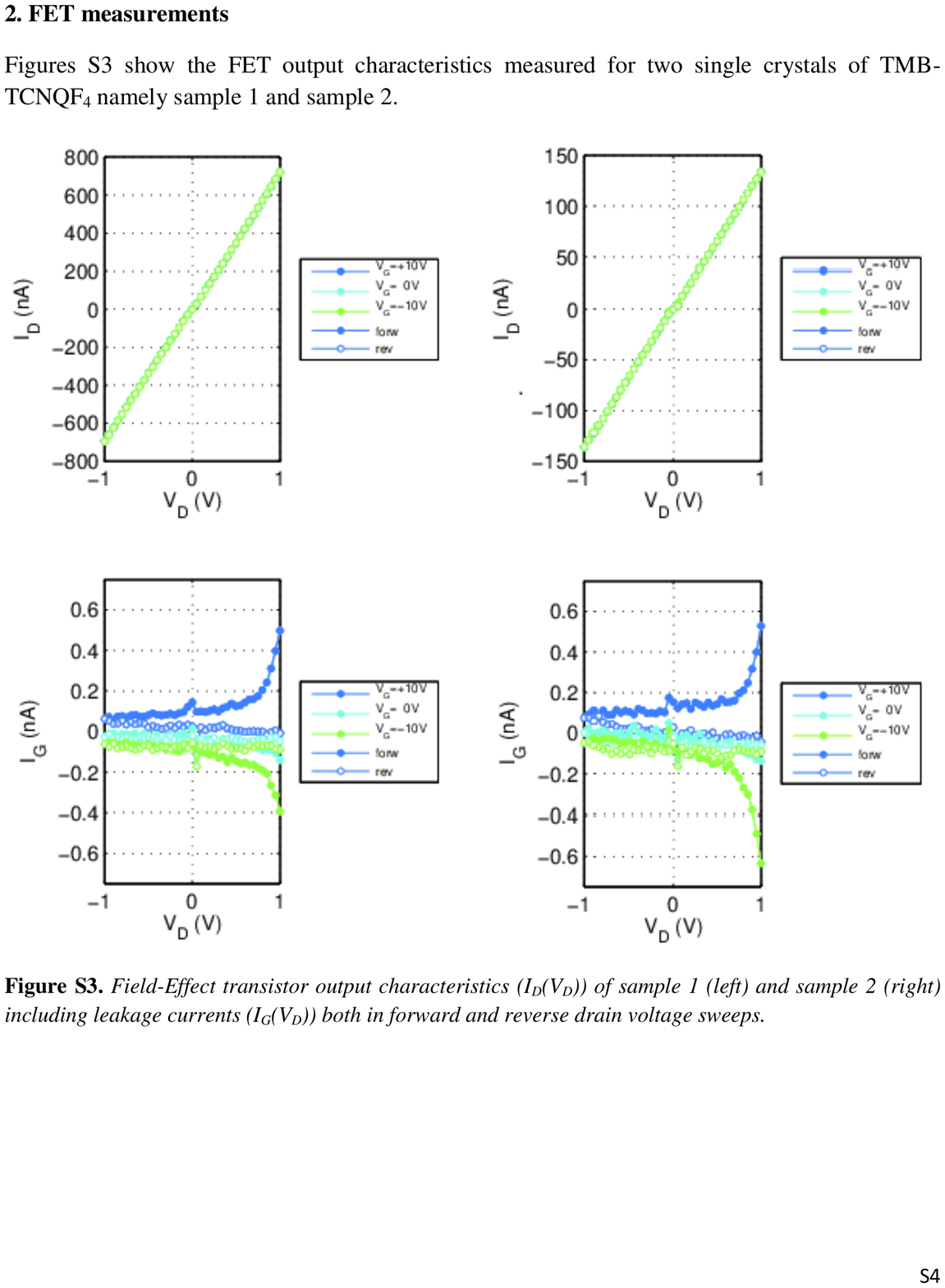}}
\end{figure}
\begin{figure}[htp]
	\centering{
		\includegraphics[scale=0.95]{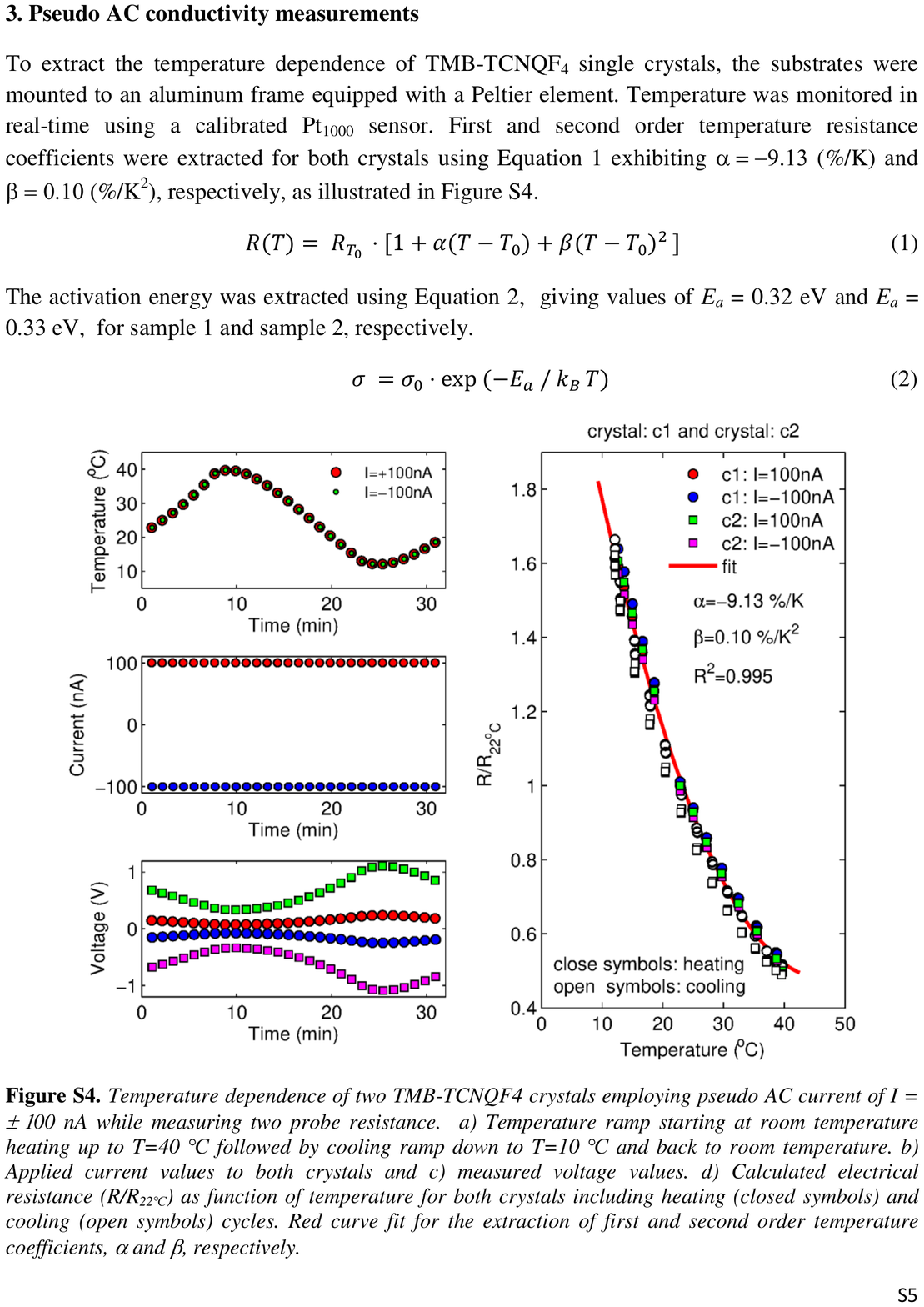}}
\end{figure}
\begin{figure}[htp]
	\centering{
		\includegraphics[scale=0.95]{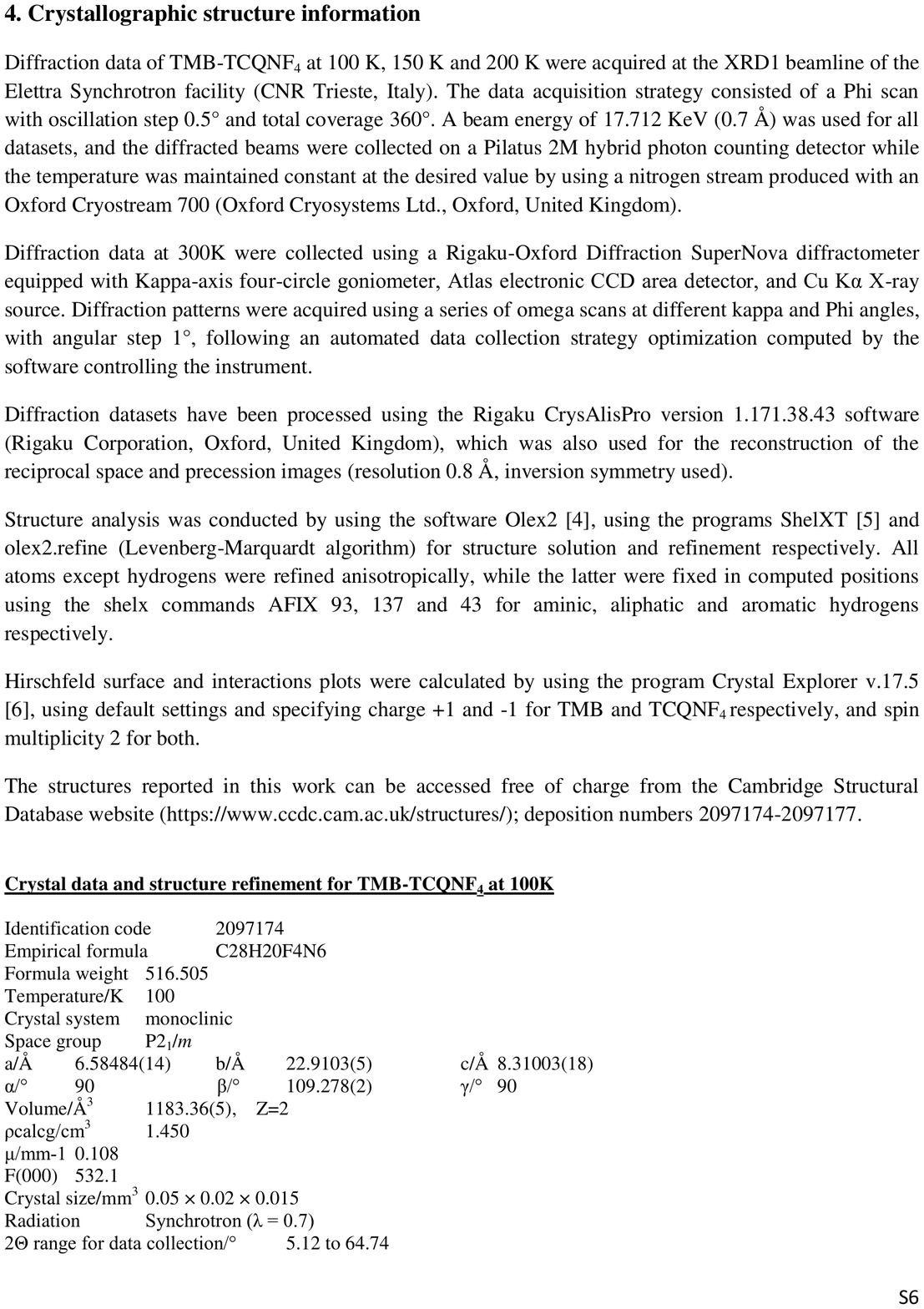}}
\end{figure}
\begin{figure}[htp]
	\centering{
		\includegraphics[scale=0.95]{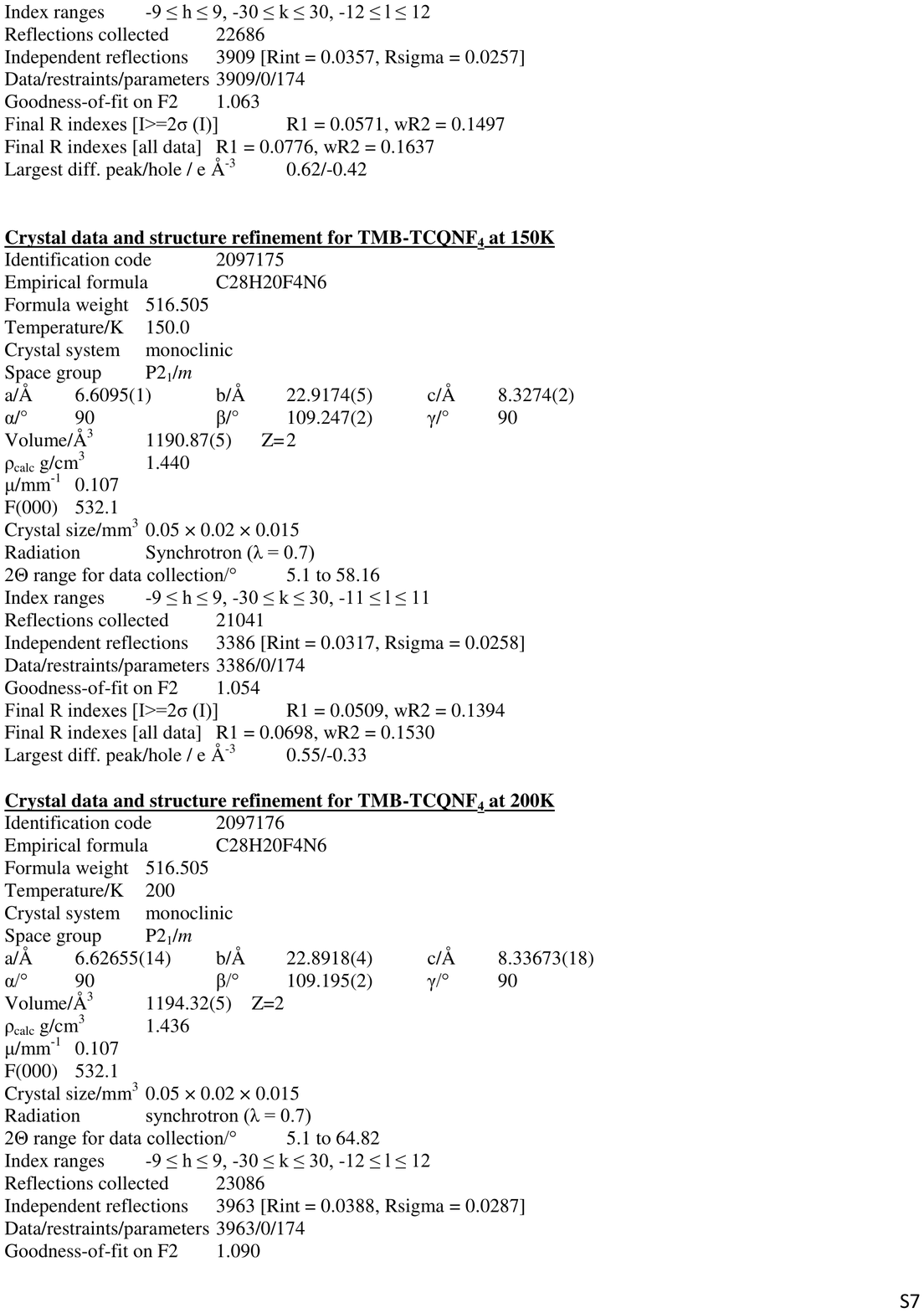}}
\end{figure}
\begin{figure}[htp]
	\centering{
		\includegraphics[scale=0.95]{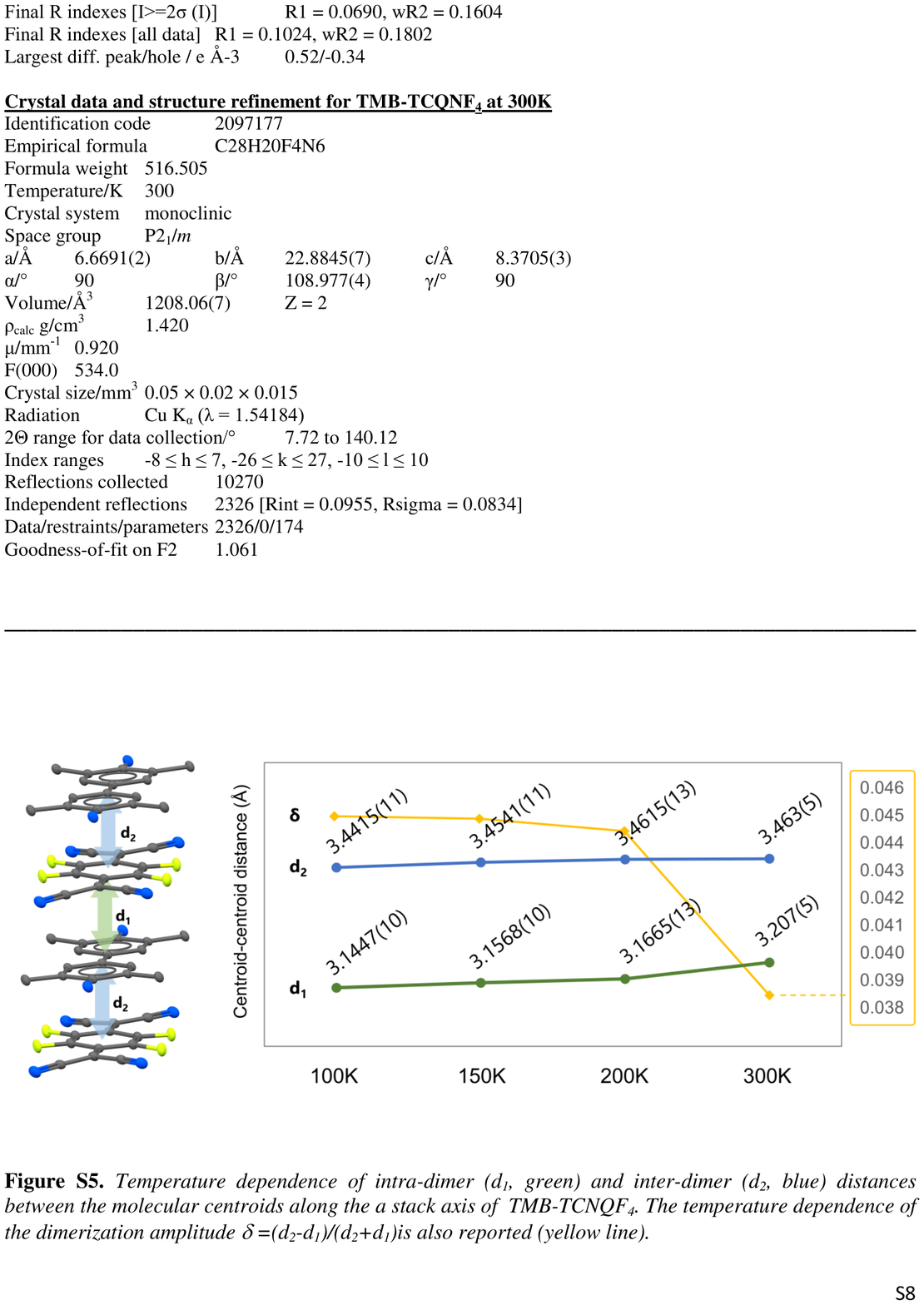}}
\end{figure}
\begin{figure}[htp]
	\centering{
		\includegraphics[scale=0.95]{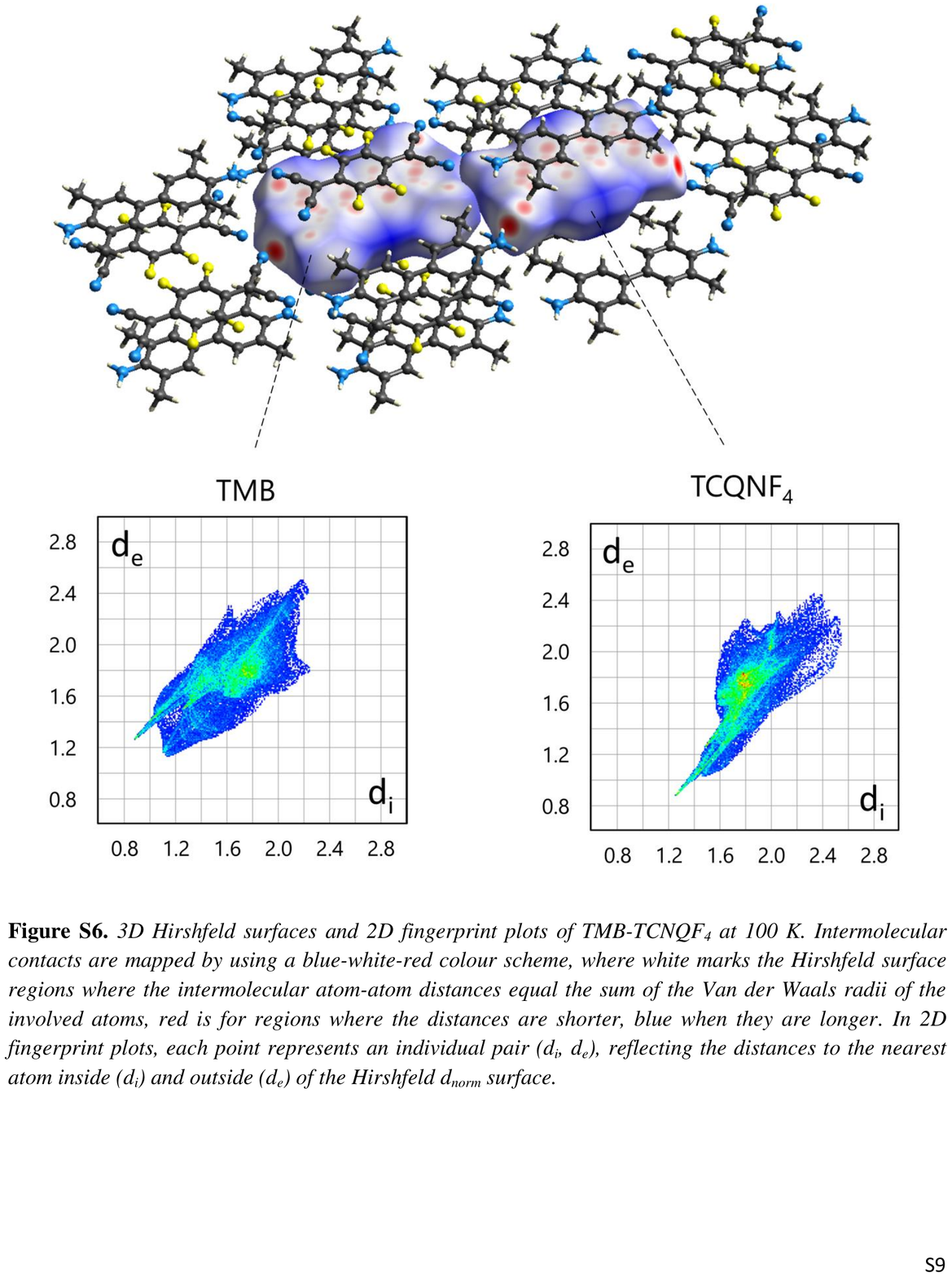}}
\end{figure}
\begin{figure}[htp]
	\centering{
		\includegraphics[scale=0.95]{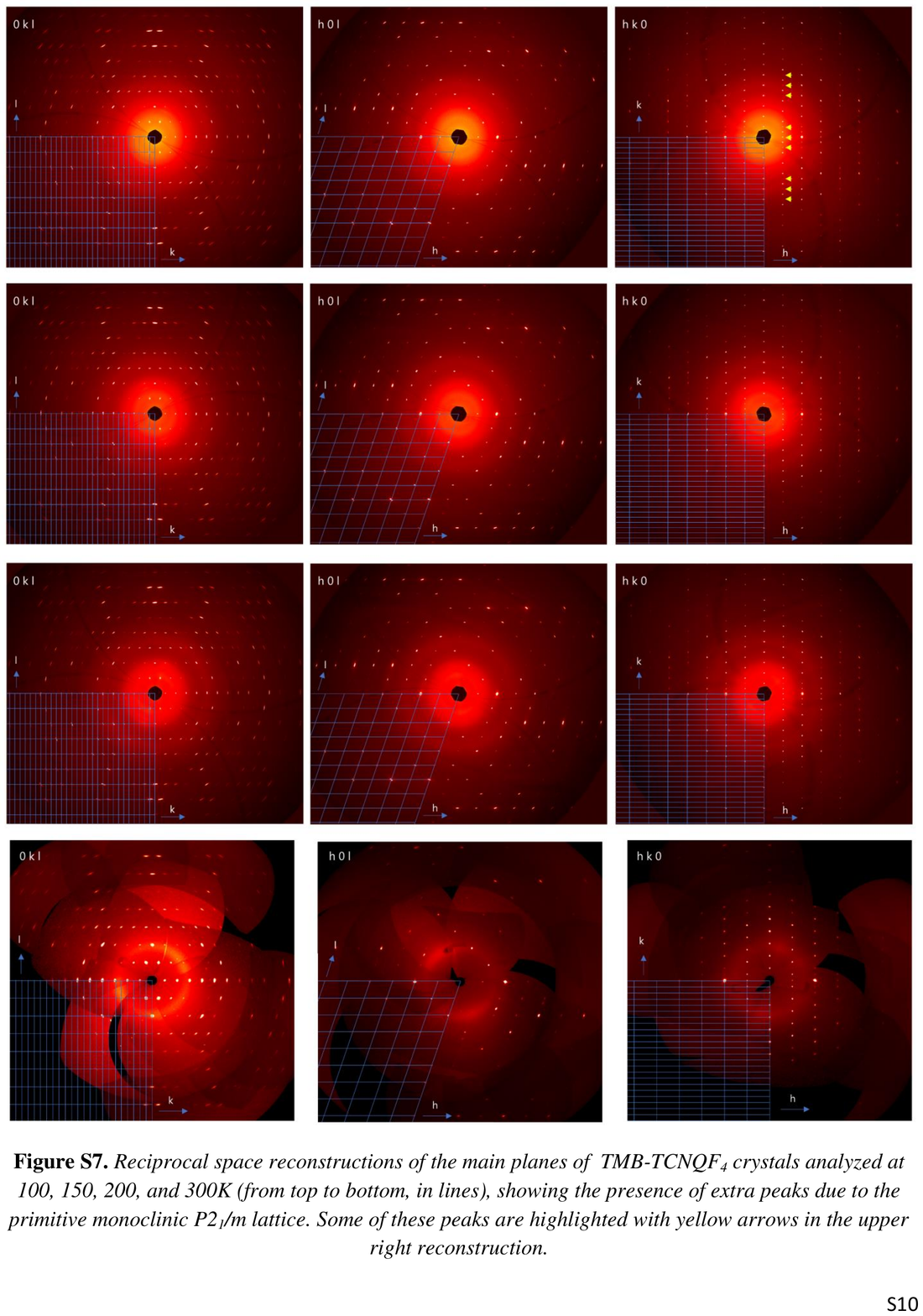}}
\end{figure}
\begin{figure}[htp]
	\centering{
		\includegraphics[scale=0.95]{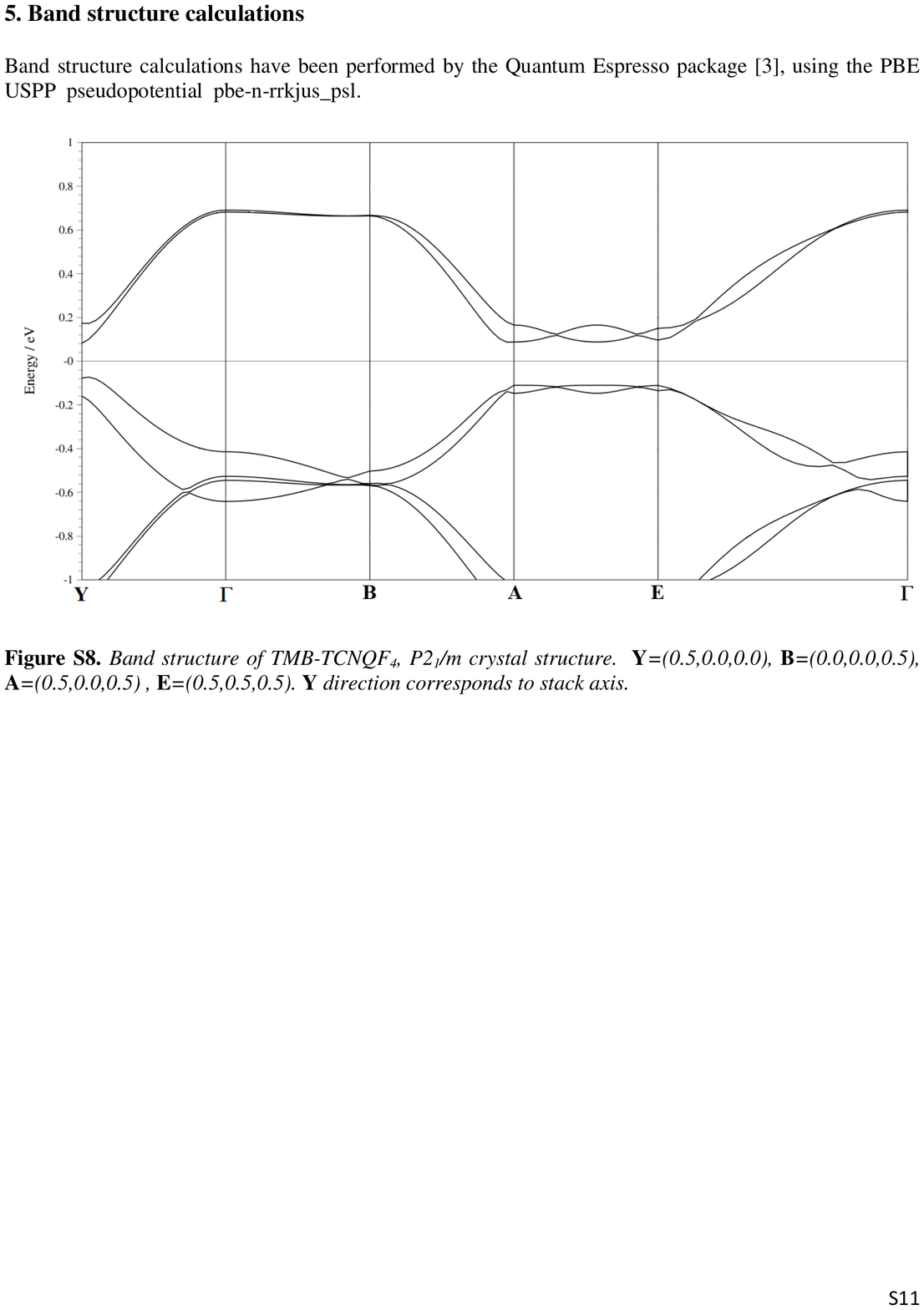}}
\end{figure}
\begin{figure}[htp]
	\centering{
		\includegraphics[scale=0.95]{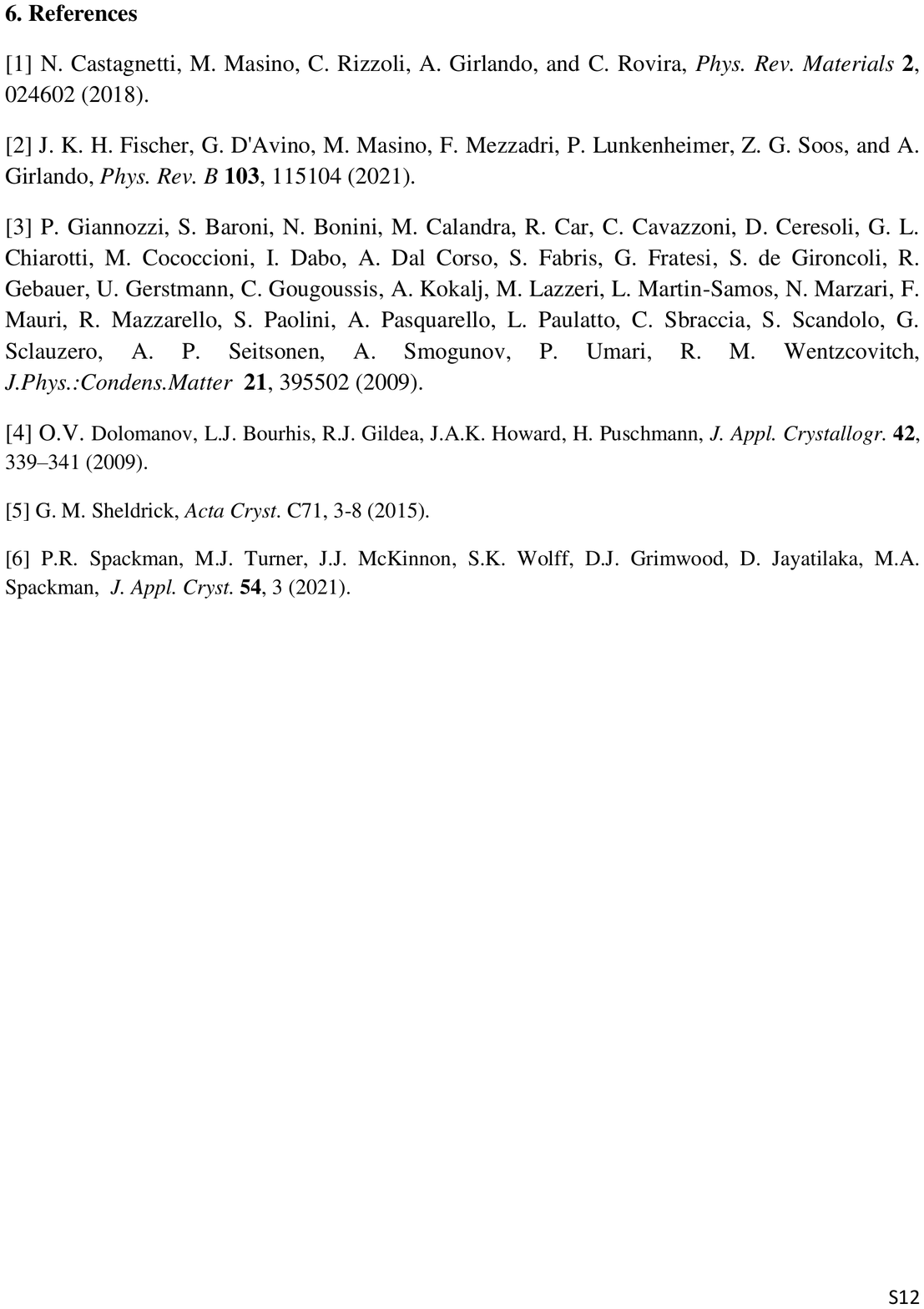}}
\end{figure}

\end{document}